\documentclass[12pt]{article}
\pdfoutput=1
\usepackage{jheppub}

\usepackage{subfigure}
\usepackage{amsmath,amssymb,amsthm,amsfonts,array,bbm,tikz}
\usepackage{graphicx}
\usepackage{color}
\usepackage[american]{babel}

\newcommand{\lp}{\left(}
\newcommand{\rp}{\right)}
\newcommand{\ti}{\widetilde}
\newcommand{\tr}{\textrm{Tr} \,}

\newcommand{\wat}{\widehat}

\newcommand{\N}{\mathcal{N}}
\newcommand{\p}{\partial}
\newcommand{\be}{\begin{equation}} \newcommand{\ee}{\end{equation}}
\newcommand{\bea}{\begin{equation} \begin{aligned}} \newcommand{\eea}{\end{aligned} \end{equation}}
\newcommand{\sh}{\text{sh}}
\newcommand{\ch}{\text{ch}}

\def\bra#1{\langle #1 | }
\def\ket#1{| #1 \rangle }

\newcommand{\cA}{\mathcal{A}}

\newcommand{\cC}{\mathcal{C}}

\newcommand{\cI}{\mathcal{I}}

\newcommand{\cN}{\mathcal{N}}

\newcommand{\cS}{\mathcal{S}}
\newcommand{\cT}{\mathcal{T}}

\newcommand{\bR}{\mathbb{R}}
\newcommand{\bZ}{\mathbb{Z}}

\def\SU{\mathrm{SU}}
\def\SL{\mathrm{SL}}

\numberwithin{equation}{section}       

\begin{document}

\begin{titlepage}

\vspace*{-2cm} 
\begin{flushright}
	{\tt  CERN-TH-2018-087} 
\end{flushright}
	
	\begin{center}
		
		\vskip .5in 
		\noindent

		{\Large \bf{Holographic duals of 3d S-fold CFTs}}
		
		\bigskip\medskip
		
		Benjamin Assel$^1$ and Alessandro Tomasiello$^2$\\
		
		\bigskip\medskip
		{\small 
			$^1$ Theory Department, CERN, CH-1211, Geneva 23, Switzerland \\
			$^2$ Dipartimento di Fisica, Universit\`a di Milano-Bicocca, I-20126 Milano, Italy\\ INFN, sezione di Milano-Bicocca		
		}
		
		\vskip .5cm 
		{\small \tt benjamin.assel@gmail.com, alessandro.tomasiello@unimib.it}
		\vskip .9cm 
		{\bf Abstract }
		\vskip .1in
\end{center}
	
\noindent
We construct non-geometric AdS$_4$ solutions of IIB string theory where the fields in overlapping patches are glued by elements of the S-duality group. We obtain them by suitable quotients of compact and non-compact geometric solutions. The quotient procedure suggests CFT duals as quiver theories with links involving the so-called $T[U(N)]$ theory.
We test the validity of the non-geometric solutions (and of our proposed holographic duality) by computing the three-sphere partition function $Z$ of the CFTs. A first class of solutions is obtained by an S-duality quotient of Janus-type non-compact solutions and is dual to 3d $\cN=4$ SCFTs; for these we manage to compute $Z$ of the dual CFT at finite $N$, and it agrees perfectly with the supergravity result in the large $N$ limit. A second class has five-branes, it is obtained by a M\"obius-like S-quotient of ordinary compact solutions and is dual to 3d $\cN=3$ SCFTs. For these, $Z$ agrees with the supergravity result if one chooses the limit carefully so that the effect of the fivebranes does not backreact on the entire geometry. Other limits suggest the existence of IIA duals.

\vfill

\end{titlepage}

\setcounter{page}{1}

\noindent\hrulefill
\tableofcontents

\noindent\hrulefill

\bigskip


\section{Introduction}
\label{sec:Intro}

The presence of dualities is one of the most striking features that sets string theory apart from other theories of gravity. It identifies configurations that would be seen as completely different by string theory's low energy supergravity approximation. In other words, the symmetry group no longer consists of diffeomorphisms alone, but also contains more exotic elements that are not geometric in nature.

This suggests the existence of solutions where the transition functions are not coordinate changes alone. Various realizations of this idea have been pursued. Perhaps the oldest and most successful is F-theory \cite{Vafa:1996xn}, a method to obtain solutions with monodromies which belong to the S-duality group $\mathrm{SL}(2,\mathbb{Z})$. These monodromies are over contractible paths, that encircle singularities which are interpreted as non-perturbative branes. Another popular example \cite{Hull:2004in} is a torus fibred over an $S^1$, with a monodromy in the T-duality group. In this case the path is non-contractible, and thus there is no singularity. Solutions of this type are called T-folds; they can be generated by T-dualizing ordinary tori in presence of NSNS flux.

It is important to explore these possibilities: they can in principle be more numerous than ordinary geometric ones, but more importantly they might evade restrictions and no-go theorems that their geometric counterparts have to satisfy. It can be difficult, however, to establish their existence, because of the very fact that they go beyond the low-energy supergravity description. If we think about an $SL(2,\bZ)$ monodromy in type IIB string theory, at the beginning and the end of the monodromy path the coupling is (in a typical situation) respectively weak and strong, 
and in the middle of the path the supergravity action cannot be trusted even after dualities. If the monodromy is over a non-contractible path, one can overcome this problem by taking the path long enough that all fields vary slowly; in this ``long-wavelength'' approximation, one expects that the two-derivative action, which is uniquely determined by supersymmetry, should suffice. 
 Such a logic is not enough in cases where the non-geometrical monodromy is over contractible paths; in that case the long-wavelength approximation will break down near the singularity encircled by the path. 

One way to confirm the validity of these constructions is to use dualities or other cross-checks. F-theory is for example often dual to M-theory, and in those cases its predictions are confirmed spectacularly. As we mentioned, T-folds can be related by T-duality to ordinary geometric backgrounds. (Sometimes a worldsheet description exists even before T-dualizing.)

Another possible way to test non-geometric solutions is to use holography. For F-theory, this has only recently started being used (for earlier discussions see \cite{Aharony:1998xz,Polchinski:2009ch}), essentially because AdS appears there less naturally than Minkowski space. AdS$_3$ solutions with non-trivial axio-dilaton were considered in \cite{Couzens:2017way,Couzens:2017nnr}. AdS$_5$ solutions were obtained in \cite{Garcia-Etxebarria:2015wns,Aharony:2016kai} by an S-quotient procedure.

In this paper, we construct ${\rm AdS}_4\times K_6$ IIB string theory solutions with monodromies in $K_6$ in the S-duality group $SL(2,\mathbb{Z})$, and we test their validity using holography. The monodromies are along non-contractible paths, so that there are no singularities encircled by them; our focus is rather on testing the limits of the long-wavelength approximation. We obtain the solutions by quotienting in various ways solutions with the local form found in \cite{D'Hoker:2007xy,D'Hoker:2007xz} (originally devised to describe the holographic dual of interfaces in ${\cal N}=4$ super-Yang--Mills). They are not related to F-theory in its present form; for example, the axio-dilaton is not holomorphic. We call them by the more general name of \textit{S-folds}. 

We consider two classes of S-folds. The first class has a monodromy given by an element $J \in SL(2,\mathbb{Z})$ with $\mathrm{Tr}J >2$ (thus in particular being a \textit{hyperbolic} element of $SL(2,\mathbb{Z})$). The geometry has the topology ${\rm AdS}_4 \times S^5 \times S^1$ with the monodromy around $S^1$. The solutions preserve $OSp(4|4)$ symmetry and are dual to 3d $\cN=4$ SCFTs. They were previously obtained in \cite{Inverso:2016eet}, which partially inspired this work, by lifting a gauged supergravity vacuum, but it can also be obtained as a quotient of a ``degenerate'' interface (or ``Janus'') solution: one where the string coupling diverges at infinity.
 This second construction points to the gauge theory dual of these $J$ solutions, since the dual of the interface has a known description 
as the infrared limit of certain 3d gauge theories involving the so-called $T[U(N)]$ theory \cite{Gaiotto:2008ak}. 

The simplest field theory in this first class consists of a single gauge group $U(N)$ with Chern--Simons coupling, which gauges the diagonal of the two $U(N)$ flavor symmetries of $T[U(N)]$ (see Figure \ref{JnQuiver}). Although the UV description for this class of 3d theories has only $\cN=3$ supersymmetry, the gravity duals indicate that the supersymmetry is enhanced to $\cN=4$ at low energies. We will support this scenario by providing an alternative description of these theories, closely related to the gravity dual solutions, as the low energy limit of a 4d $\cN=4$ $U(N)$ SYM theory on a circle whose coupling varies and has a $J$ monodromy around the circle, while preserving 3d $\cN=4$ supersymmetry. Although the resulting theory is not fully Lagrangian, assembling known ingredients  we can compute its three-sphere partition function $Z$. Remarkably, this turns out to be a Gaussian integral, which we manage to solve fully, even at finite $N$, with Fermi gas techniques. In the large $N$ limit, this result agrees with the result one obtains from the supergravity solution: $F\equiv -\mathrm{ln}Z \sim f(J) N^2$, with a coefficient that depends on $J$ and that is reproduced exactly. This provides a strong confirmation of the existence of this class of S-folds.\footnote{The same 4d SYM setup with the Janus configuration with $J$-monodromy and the resulting 3d low-energy theories were previously studied in \cite{Ganor:2014pha} for abelian gauge groups (single D3 brane), in which case the 3d theory reduces to a Chern--Simons quiver. The identity \eqref{eq:detQ} that we use in our holographic test appears already in this work, and is interpreted there as an equality of Hilbert space dimensions. We provide an alternative derivation of it.}

Emboldened by this success, we then investigate a second, more challenging class, where brane singularities are also included. Again the $SL(2,\mathbb{Z})$ monodromies are over a non-contractible path, so that there are no singularities that can be interpreted as seven-branes as in F-theory. But the class of local solutions in  \cite{D'Hoker:2007xy,D'Hoker:2007xz} allows to include NS5-branes and D5-branes wrapping various $S^2$ submanifolds; in fact for fully geometrical global solutions these have to be included \cite{Assel:2011xz, Assel:2012cj}. In an S-fold this is not necessarily the case, as the above-mentioned $J$ class demonstrates; but branes complicate the applicability of the supergravity approximation in interesting ways. We obtain solutions in this second class by quotienting a geometrical solution by an involution that mixes a geometrical and an $SL(2,\mathbb{Z})$ action. 
The original geometric solution has still an $S^5\times S^1$ internal space with five-brane singularities wrapping $S^2$s; the geometrical part of the quotient acts as a rotation (an order four involution) on the $S^5$ and as a shift on the $S^1$. We call the resulting solutions \textit{S-flip solutions}. The monodromy of the resulting solution is the S-duality element $S=\left(\begin{smallmatrix}
0 & -1 \\ 1 & 0
\end{smallmatrix}\right) \in SL(2,\mathbb{Z})$ and is along the non-contractible $S^1$ circle. Part of the supersymmetry of the original geometrical solution is broken by the quotient. The preserved superconformal algebra is $OSp(3|4)$  and the dual SCFT has only $\cN=3$ supersymmetry. We did not find S-fold solutions in this class with a monodromy by another $SL(2,\mathbb{Z})$ element.

The field theory duals for this second class are necklace quivers where one link is not an ordinary bifundamental hypermultiplet but rather a $T[U(N)]$ link: namely, a $T[U(N)]$ theory whose two $U(N)$ flavor symmetries are gauged by two neighboring gauge groups. Each theory has in fact several dual realizations. The simplest example is a theory with two $U(N)$ nodes connected by a bifundamental hypermultiplet and a $T[U(N)]$ link (see Figure \ref{halfABJM}), that we call ``half-ABJM'', because it comes about by a quotient of a solution which is holographic dual \cite{Assel:2012cj} to the ABJM theory \cite{Aharony:2008ug}. It has a necklace generalization with $\hat M+1$ gauge groups, where one of the links is a $T[U(N)]$ link, as described above, and one of the gauge groups has $M$ fundamental hypermultiplets. For these theories we compute the three-sphere partition function in terms of a matrix model; it depends only on $K\equiv M + \hat M$. The long-wavelength approximation suggests making the monodromy path long; this requires $N\ll K^2$. However, the branes now introduce a local region where the supergravity action changes rapidly. Such a region is of course present in any solution with branes; past experience with holography suggests simply imposing that this region does not eat up the entire geometry, which imposes $N \gtrsim K$. We are able to evaluate the limit of the matrix model at the lower end of this window, getting $F \sim \frac12 N^2 \mathrm{ln} \, N$; again we find agreement with the supergravity results. 

One might be curious about what happens if one pushes the limits of the long-wavelength approximation. In the limit $N\gg K^2$, the monodromy path is small and the branes are effectively smeared. Their backreaction is felt all over the geometry; this suggests that the supergravity approximation might break down.
On the field theory side, for $K=1$ we can evaluate the matrix model analytically at large $N$: this case corresponds to the half-ABJM theory mentioned earlier.  A computation rather similar to \cite{Jafferis:2011zi} produces a behavior that is unlike what the two-derivative supergravity action would predict, as expected. A surprise is that in this limit $F \sim N^{5/3}$, which happens to be the same as models with massive IIA holographic duals. This might suggest a dual IIA description of our solution in this limit, which at present is not obvious. 

The rest of the paper is organized in two main sections, each devoted to the study of one of the two classes of solutions we just described.  In section \ref{sec:Janus} we consider Janus-type S-folds; after describing the general idea, we focus in section \ref{ssec:JanusfoldSolutions} on an example corresponding to a particularly simple element $J\in SL(2,\mathbb{Z})$; we describe its field theory dual in section \ref{ssec:JnfoldSCFTs}. In section \ref{ssec:TestJanus} we then check that the three-sphere partition function computed with field theory and supergravity methods indeed match. In section \ref{ssec:GeneralJanusfold} we then consider the generalization to any element $J\in SL(2,\mathbb{Z})$ with $\mathrm{Tr} J>2$. We then proceed in section \ref{sec:SfoldQuivers} to the second class, that of S-flip  solutions. Again we first illustrate the class with an example, considered from the field theory and supergravity points of view in sections \ref{ssec:HalfABJM} and \ref{ssec:DualBkgrd} respectively. We then go on to construct more general examples in this class in section \ref{ssec:GenSol}, before performing a holographic check in section \ref{ssec:FreeEnergy}.


\section{Janus S-fold solutions and SCFTs}
\label{sec:Janus}

The 10d type IIB supergravity solutions that we construct in this paper are obtained by a certain S-folding procedure applied to a class of solutions whose local form was found in \cite{D'Hoker:2007xy,D'Hoker:2007xz}. These solutions describe an ${\rm AdS}_4\times S^2\times S^2\times \Sigma$ geometry, where $\Sigma$ is a Riemann surface, which admit 16 Killing spinors and are dual to SCFTs with 3d $\cN=4$ supersymmetry. The $SU(2)\times\SU(2)$ R-symmetry is reflected in the isometries of the two two-spheres. 
Global solutions in this class were proposed in \cite{Assel:2011xz, Assel:2012cj} for $\Sigma$ having the topology of a disk or an annulus with five-brane singularities on the boundary\footnote{The points on the boundary of $\Sigma$ are still interior points of the geometry, due to the vanishing of a two-sphere with appropriate scaling.}
 as gravity duals of a class of 3d $\cN=4$ linear and circular quivers. Other solutions, which were in fact the initial solutions found in \cite{D'Hoker:2007xy,D'Hoker:2007xz}, are such that $\Sigma$ is an infinite (non-compact) strip with two asymptotic ${\rm AdS}_5\times S^5$ regions and are holographic duals of 3d $\cN=4$ defect SCFTs in 4d $\cN=4$ SYM. The simplest example in this class is the Janus solution which we will discuss shortly.

The solutions are elegantly parametrized by two real harmonic functions on $\Sigma$, denoted $h_1,h_2$, which obey some boundary conditions on the boundary of $\Sigma$.\footnote{Alternatively the solutions can be parametrized by two holomorphic functions $\cA_1,\cA_2$ on $\Sigma$ with certain boundary conditions, which have shift ambiguities.} A summary of the local supergravity solution in terms of $h_1,h_2$ is given in Appendix \ref{app:LocalSol}. In known solutions $\Sigma$ is an infinite strip or an annulus and is parametrized  by a complex coordinate $z =x+iy$, with $x$ periodic for the annulus and $y\in [0,\frac{\pi}{2}]$. On the upper boundary $y=\frac{\pi}{2}$ one two-sphere shrinks to zero size, while on the lower boundary $y=0$ the other two-sphere shrinks to zero size.

\medskip

In this section we construct supergravity solutions by S-folding a special Janus solution, we propose a holographic dual 3d $\cN=4$ SCFT and perform a non-trivial test of the holographic duality. The simplest S-fold supergravity solutions that we find reproduce solutions described in \cite{Inverso:2016eet}.

\subsection{Supergravity solutions}
\label{ssec:JanusfoldSolutions}

\begin{figure}[h!]
\centering
\includegraphics[scale=0.8]{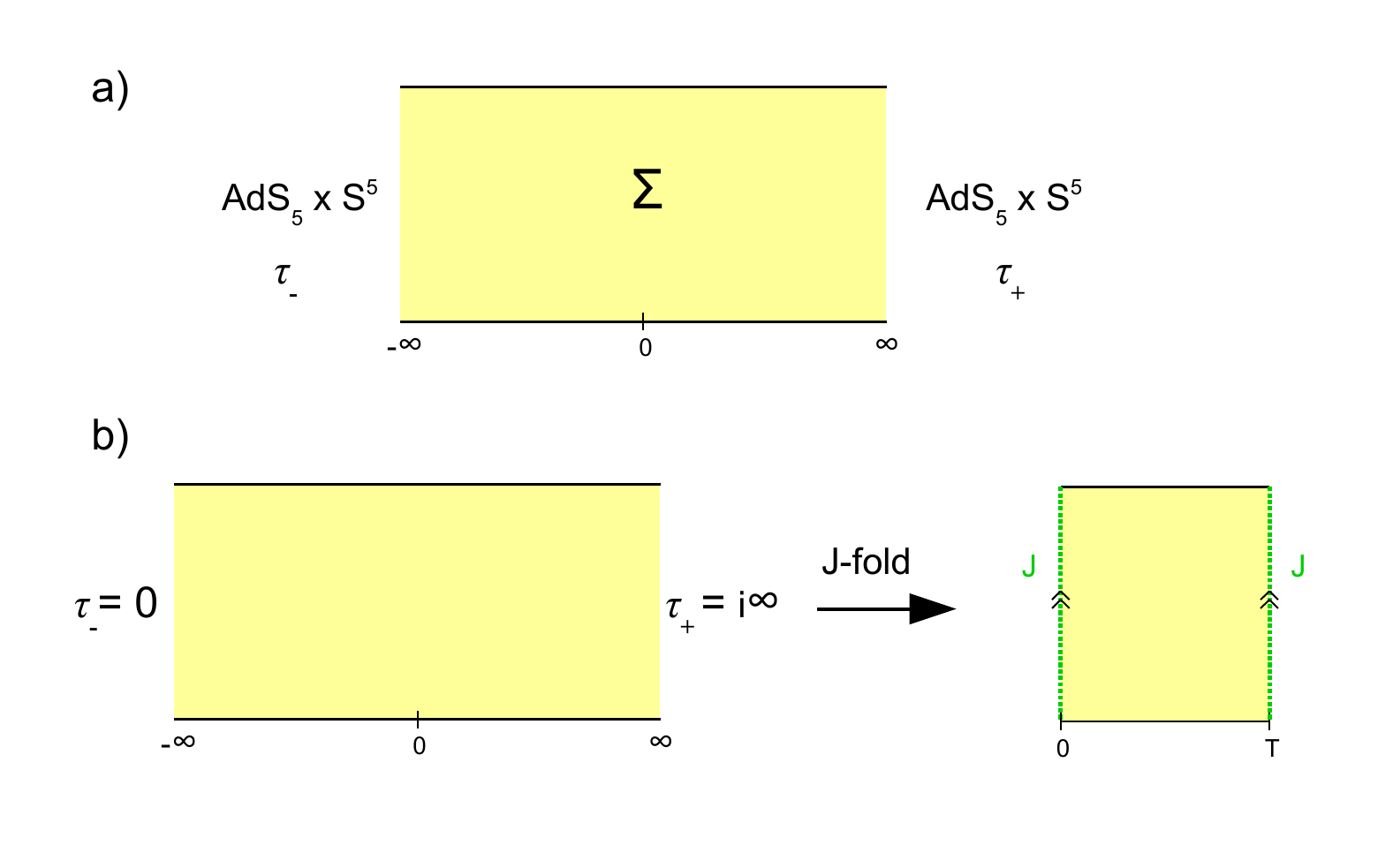} 
\vskip -1cm
\caption{\footnotesize a) Picture of $\Sigma$ (yellow strip) for the Janus solution with asymptotic axio-dilaton values $\tau_\pm$. b) $J$-folding of the extremal Janus solution. This involves an $SL(2,\bR)$ transformation before taking the $J$-fold quotient. The resulting solution has a cut (green) with $J$ monodromy.}
\label{Janusfold}
\end{figure}

The supersymmetric Janus supergravity solution is the holographic dual background to the Janus interface theory in 4d $\cN=4$ $U(N)$ SYM. The simplest Janus interface is characterized by having varying gauge coupling $\tau(x') = \frac{4\pi i}{g(x')^2}$ along a space direction $x'$, while preserving 3d $\cN=4$ supersymmetry. It was introduced in \cite{DHoker:2006qeo} and generalized in \cite{Gaiotto:2008sd}.\footnote{See also \cite{Kim:2008dj,Kim:2009wv} for other studies of the supersymmetric Janus theory.} 
The holographic dual background corresponds to a solution with $\Sigma$ an infinite strip, which is shown in Figure \ref{Janusfold}. Their ten-dimensional topology is that of AdS$_4\times S^5 \times \bR$.
The harmonic functions, as re-expressed in \cite{Assel:2011xz}, are
\bea
h_1(z,\bar z) &= - i \alpha \sinh(z-\beta) + \, c.c. \cr
h_2(z,\bar z) &= \hat\alpha \cosh(z-\hat\beta) + \, c.c. \,,
\label{JanusSol}
\eea
with real parameters $\alpha,\hat\alpha,\beta,\hat\beta$ and we choose $\alpha,\hat\alpha > 0$.\footnote{Other choices are obtained by charge conjugations.}
The complex coordinates $z=x+iy$ spans the infinite strip with 
\begin{equation}
	-\infty < x < +\infty\quad \text{and} \quad 0 \le y \le \frac{\pi}{2}\,.
\end{equation}

The asymptotic regions $x \to \pm \infty$ are ${\rm AdS}_5\times S^5$ spaces with identical radii $L$, but with different dilaton values $e^{2\phi^\pm}$ ($\tau_{\pm} = ie^{-2\phi^\pm}$ in Figure \ref{Janusfold}-a) ,
\bea
L^4 = 16 \alpha\hat\alpha \cosh(\beta - \hat\beta) \,, \quad e^{2\phi^\pm} = \frac{\hat\alpha}{\alpha} e^{\pm(\beta - \hat\beta)} \,.
\eea
The geometry has a 5-cycle with the topology of a 5-sphere $\cC_5 = \cI\times S^2_1 \times S^2_2 \cong S^5$, where $\cI$ is an interval going from the upper to the lower boundary of $\Sigma$ and supporting $N$ units of 5-form flux, with
\be
N = \frac{1}{(4\pi^2\alpha')^2}\int_{\cC_5} F_5 = \frac{L^4}{2^6\pi} \,,
\ee
in the convention $\alpha'=4$. The 5-form flux is independent of the position of $\cC_5$ along $x$ and therefore spans the whole geometry.
If $\beta=\hat\beta$ the solution is globally ${\rm AdS}_5\times S^5$.
A change of variables $x \to x + c$ can be used to set $\hat\beta=-\beta$ if desired. The solution has then three parameters.  

We now consider a degenerate limit of the Janus solution with $\beta=-\hat\beta \to -\infty$, $\alpha,\hat\alpha \to 0$ and $\alpha e^{-\beta}= c > 0$, $\hat\alpha e^{\hat\beta}= \hat c > 0$ fixed, leading to
\bea
\label{JanusExtremal}
h_1(z,\bar z) &= \frac{c}{2i} \, ( e^{z} - e^{\bar z})  \cr
h_2(z,\bar z) &= \frac{\hat c}{2} \, (e^{-z} + e^{-\bar z}) \,.
\eea
The asymptotic regions in this limit have ${\rm AdS}_5$ radius $L^4 = 8 c \hat c$ and diverging dilaton $\phi^+ = -\infty$, $\phi^- = +\infty$, which is why we call it a degenerate limit.\footnote{This solution can also be reached from the solution dual to the $T[U(N)]$ theory, by sending the five-brane stacks to infinity, as was studied in \cite{Assel:2012cp} and in \cite{Lozano:2016wrs} where this extremal background appeared from a non-abelian T-duality action on $AdS_4$ IIA solutions.} Constant shifts in $x$ allows to set $c=\hat c$, so that there is really only a one-parameter family of such degenerate solutions (with discrete parameter $N$). This solution was already found in \cite{Inverso:2016eet} from a different construction.

One nice property of this limit is that the dependence on $x$ becomes very simple, with all fields being independent of $x$ (in particular the metric) except for the dilaton and the three-form fields\footnote{The explicit expression for the non-trivial 5-form can be worked out from the formula of \cite{D'Hoker:2007xy,D'Hoker:2007xz} in terms of $h_1,h_2$.} 
\bea
ds^2 &= (c \hat c) ^{1/2} \Big[ 2^{\frac 14} (7- \cos(4y))^{\frac 14} ds^2_{{\rm AdS}_4} + 2^{\frac{3}{2}}\frac{(2+\cos(2y))^{\frac 14}}{(2-\cos(2y))^{\frac 34}} \sin(y)^2 ds^2_{S^2_{(1)}} \cr
& \quad + 2^{\frac{3}{2}}\frac{(2-\cos(2y))^{\frac 14}}{(2+\cos(2y))^{\frac 34}} \cos(y)^2 ds^2_{S^2_{(2)}} + 2^{\frac 54} (7- \cos(4y))^{\frac 14} (dx^2 + dy^2) \Big]  \,, \cr
e^{2\phi} &= \frac{\hat c}{c} \lp\frac{2+\cos(2y)}{2-\cos(2y)}\rp^{\frac 12} e^{-2x} \,, \cr
H_3 &=  \omega_{S^2_{(1)}} \wedge db_1 \,, \quad b_1 =  8 \hat c \, \frac{\sin(y)^3}{2-\cos(2y)} e^{-x}  \,,\cr
F_3 &= \omega_{S^2_{(2)}} \wedge db_2 \,, \quad b_2 = -8 c \, \frac{\cos(y)^3}{2+\cos(2y)} e^{x}  \,,
\label{ISTfields}
\eea
with $ds^2_{{\rm AdS}_4}$, $ds^2_{S^2_{(i)}}$, the unit radius metrics on AdS$_4$ and the two-spheres respectively. We have $N =  \frac{c\hat c}{8\pi}$, while $\frac{\hat c}{c}$ is a free (unphysical) parameter.
These fields are those transforming under $SL(2,\bZ)$ type IIB (gauge) symmetry (see Appendix \ref{app:LocalSol}) and one may look for a symmetry of the solution under the combined action of a translation along $x$ and an $SL(2,\bZ)$ transformation $J$. If such a symmetry exists we can quotient the solution by its action and produce an S-fold solution with compact $x$ direction and $J$ monodromy. Unfortunately no such symmetry exists in the above solution. 
However one may generate {\it new} solutions by applying $SL(2,\bR)$ transformations to the degenerate solution \eqref{JanusExtremal}.
A new solution obtained that way may then admit the desired symmetry, and thus would allow to define an S-fold solution.

In order for this scenario to work their must exist $M\in \SL(2,\bR)$, $J\in \SL(2,\bZ)$ and $T>0$, such that the $M$-transformed extremal Janus solution is invariant by a translation by $T$ along $x$ combined with the action of $J$. For the $SL(2,\bR)$-doublet $(H_3,F_3)$, this translates into the condition
\be
M^{-1}J^{-1}M = \lp
\begin{array}{cc}
e^{-T} & 0 \cr
0 & e^{T} 
\end{array}\rp
\,.
\label{SfoldCondition}
\ee
The simplest solutions are found by taking
\be 
J  = 
\lp
\begin{array}{cc}
n & 1 \\
-1 & 0 
\end{array}\rp  := J_n  \,.
\ee
A short analysis shows that there is a solution for $n > 2$,\footnote{To be precise there is a continuous family of solutions $M(\lambda)$, $\lambda \in \bR^\ast$ for a given $n$, which implement the scalings of the extremal Janus solutions $(c,\hat c) \to (\lambda c,\lambda^{-1}\hat c)$. They all correspond to the same supergravity solution since this rescaling is equivalent to a translation along $x$.}
\bea\label{JnParam}
& n = e^{T}+ e^{-T} \quad \leftrightarrow \quad T = \ln\lp \frac 12 \lp n + \sqrt{n^2 - 4} \rp \rp \,, \cr
& M =  \lp
\begin{array}{cc}
\frac{1}{1+e^{-T}} & \frac{1}{1-e^{T}} \cr
\frac{-1}{1+e^{T}} & \frac{1}{1-e^{-T}}  
\end{array}\rp \,.
\eea
One can check that the transformed axio-dilaton $\tau' = M.\tau$ obeys $J_n.\tau'(x+T) = \tau'(x)$.  The $M$-transformed solution is thus invariant under the action of $\cT$ which is the combination of a translation by $T$ along $x$ and the $SL(2,\bZ)$ transformation $J_n$. It allows to define the quotient of the $M$-transformed solution by $\cT$ which we call the {\it  $J_n$-fold solution}. The resulting topology after the $\bZ$-quotient is that of AdS$_4\times S^5 \times S^1$.

Let us describe explicitly the $M$-transformed solution whose quotient defines the $J_n$-fold solution. The metric and five-form are that of the extremal Janus solution and are constructed from $h_1,h_2$ in \eqref{JanusExtremal} (see Appendix \ref{app:LocalSol}). The axio-dilaton $\tau'= \chi' + i \, e^{-2\phi'}$ and the three-forms are
\bea
& \tau' = \frac{(1-e^{-T})^{-1} i \, e^{-2\phi} - (1 + e^{T})^{-1}}{(1-e^{T})^{-1} i \, e^{-2\phi} +(1 + e^{-T})^{-1}} \,, \cr
& \binom{H'_3}{F'_3} = M \binom{H_3}{F_3}\,, 
\label{Jnfoldsolution}
\eea
with $e^{2\phi}, H_3$ and $F_3$ constructed from $h_1,h_2$ in \eqref{JanusExtremal}.
In the $J_n$-fold solution we have spatial periodicity $(x,y) \sim (x+T,y)$, so that $\Sigma$ is topologically an annulus, and the gluing conditions at $x=T\sim 0$ involve a $J_n$ transformation of the fields  (i.e. $J_n$-twisted boundary conditions or $J_n$ monodromy). This $J_n$ folding procedure is schematically depicted in Figure \ref{Janusfold}-b for a generic $J$-folding.
This reproduces the S-fold solutions mentioned in \cite{Inverso:2016eet}.

One can also construct $J_n^{-1}$-fold solutions in a similar way. In general one can obtain a $J^{-1}$-fold solution from a $J$-fold solution by taking the matrix $M$ to be $M_{J^{-1}}(T) = M_{J}(-T)$. Finally, global $SL(2,\bZ)$ actions map a $J$-fold solution to a conjugate $J'^{-1}JJ'$-fold solution. Using such manipulations one can construct a close cousin to the $J_n$-fold solution: a $\bar J_n$-fold solution with $\bar J_n =-J_{-n}$. 

 Before moving to other $J$-fold solutions we first study the holography of the $J_n$-fold solutions.

\subsection{CFT duals}
\label{ssec:JnfoldSCFTs}

We now describe the 3d $\N=4$ field theories dual of the $J_n$-fold supergravity solutions.
To start with, the Janus supergravity solution \eqref{JanusSol} is dual to the Janus interface CFT \cite{DHoker:2006qeo,Gaiotto:2008sd}, which is the 4d $\N=4$ SYM theory with complex coupling $\tau$ jumping across a 3d interface from a value $\tau^-= i e^{-2\phi^-}$ to a value $\tau^+ = i e^{-2\phi^+}$.\footnote{We use abusively the same name $\tau$ to denote both the SYM coupling and the type IIB axio-dilaton.} It is useful to think about this theory as the infrared limit of 4d $\N=4$ SYM with a smoothly varying coupling along a space direction parametrized by $x'$, with $\lim_{x' \to \pm\infty} \tau(x') = \tau^\pm$. The exact profile of $\tau(x')$ is irrelevant in the low-energy limit. The configuration is constructed so that it preserves 3d $\N=4$ supersymmetry.

Sine the $J_n$ solution is a circle compactification of the (extremal) Janus solution with a $J_n$ twist, it is natural to conjecture that their 3d CFT duals are obtained as the low-energy limit of a circle compactification of the Janus 4d theory with $J_n$ twisted boundary conditions.
We thus look for a Janus configuration which is periodic up to a $J_n$ transformation, namely an $\N=4$ Janus solution with $J_n.\tau(x'+T')=\tau(x')$, for some $T'>0$.
The $\cN=4$ preserving profiles of $\tau(x')$ are given by \cite{Gaiotto:2008sd}
\be
\tau(x') = a +  D e^{i\psi(x')} \,,
\label{JanusProfiles}
\ee
where $a\in \bR$ and $D\in \bR_{\ge 0}$ are arbitrary constants and $\psi(x)$ is any function such that $\tau(x')$ stays in the upper half plane. This means that the trajectories $\tau(x')$ must stay on a circle in the upper half-plane. We must look for such profiles which satisfy $\frac{-1}{\tau(x'+T')+n}=\tau(x')$. 
We already have candidate solutions which do satisfy this equation. These are simply the $\tau'(x,y)$ profiles of the axio-dilaton in the $J_n$-fold solutions \eqref{Jnfoldsolution} for any fixed value of $y$, which satisfy $J_n.\tau'(x+T,y)=\tau'(x,y)$. So we can try to pick the varying SYM coupling to be
\be
\tau(x') = \frac{(1-e^{-T})^{-1} i \, \lambda \, e^{2x'} - (1 + e^{T})^{-1}}{(1-e^{T})^{-1} i \, \lambda \, e^{2x'} +(1 + e^{-T})^{-1}} \,.
\label{tauprime}
\ee
with any $\lambda \in \bR^\ast$ and $T'=T = \ln [\frac 12 (n + \sqrt{n^2-4}) ]$. For this to be a solution to our problem we must show that it can be written in the form \eqref{JanusProfiles}. We find that it is indeed the case, with
\be
a = - \frac 12 \left(e^{T}+ e^{-T}\right)  \,, \quad D = \frac 12 \left(e^{T} - e^{-T}\right)  \,, \quad e^{i\psi(x')} = \frac{1 - e^{T} - i \lambda (1+ e^{-T}) e^{2x'}}{1 - e^{T} + i \lambda (1+ e^{-T}) e^{2x'}} \,.
\ee
With $\lambda < 0$, $\tau(x')$ is in the upper half plane. The parameter $\lambda$ is here again irrelevant since it can be fixed to one (or minus one) by a shift in $x'$. A larger class of solutions is obtained by replacing $\lambda$ with any negative periodic function $\lambda(x')$ with period $T'$. One can show that this covers all solutions to the problem.
The solutions \eqref{tauprime} are somewhat degenerate Janus configurations in the sense that the coupling $\tau'(x)$ becomes real at $x'=\pm \infty$, corresponding to infinite Yang--Mills coupling. Of course this is completely analogous to the gravity construction.

These specific 4d Janus theories admit a quotient by the combined action of a translation by $T'$ and a $J_n$ S-duality action. They lead to a 4d $\N=4$ theory on a circle with $J_n$-twisted boundary conditions, which preserves 3d $\N=4$ supersymmetry. The infrared limit of such a configuration is a 3d $\N=4$ SCFT which we propose as the CFT dual of the $J_n$-fold supergravity solutions. On physical grounds, we do not expect the 3d limit to depend on the explicit choice of profile $\lambda(x')$ along the $x'$ circle. The infrared limit should only depend on the monodromy $J_n$.
The resulting 3d $\N=4$  SCFTs are thus labeled by $n$ and the rank $N$ only, matching the gravity data. We will call these 3d CFTs {\it $J_n$ theories}. This construction of 3d theories from Janus configurations on a circle with duality-twisted boundary conditions was already discussed in \cite{Ganor:2014pha}.\footnote{A related construction of duality surface defects in $\N=4$ SYM with $SL(2,\bZ)$ monodromies was studied in \cite{Martucci:2014ema,Assel:2016wcr} (see also \cite{Gadde:2014wma}).}

\medskip

The twist of the boundary conditions by an $SL(2,\bZ)$ element $J$ has no (known) description in terms of gluing conditions on local fields.\footnote{ Except for $J=T^k$ duality interfaces, which have all fields continuous across the interface and a 3d Chern--Simons term at level $k$ on the 3d interface.} The 3d interface is only the boundary of a patch (of a non-trivial $SL(2,\bZ)$ bundle) and does not carry local degrees of freedom and we are free to move the location of the interface without affecting the theory.

It is possible to obtain a quasi-Lagrangian UV description of the $J_n$ theories. In the description as 4d $\cN=4$ SYM on a circle with $J_n$ twisted boundary conditions, we can choose a convenient profile by adjusting the periodic function $\lambda(x')$ in \eqref{tauprime}, since this should not affect the 3d limit.
In particular we can tune the $\tau$ profile until it becomes almost constant along $x'$ with the variation confined to a tiny region close to the jump at $x'=T \sim 0$. We obtain a  configuration which can be described in the UV as 4d $\cN=4$ SYM on a circle coupled to a 3d theory with a quasi-Lagrangian description.
Such constructions were studied in \cite{Gaiotto:2008ak}, and the 3d theory associated to the monodromy $J_n = - ST^n$ is the $T[U(N)]$ theory with a non-abelian Chern--Simons term at level $n$ for one of its two $U(N)$ flavor groups. The 3d Chern--Simons term preserves only $\cN=3$ supersymmetry but, since the Janus setup preserves $\cN=4$, the supersymmetry must be enhanced at low-energies.

The $T[U(N)]$ theory was introduced in \cite{Gaiotto:2008ak} as the IR SCFT of a linear quiver theory with a chain of unitary gauge nodes of increasing rank, from $U(1)$ to $U(N-1)$, and with $N$ fundamental hypermultiplets in the $U(N-1)$ node. The UV global symmetry $U(1)^{N-1}\times SU(N)$ is enhanced to $SU(N)\times SU(N)$ in the infrared SCFT. In addition, in the $T[U(N)]$ theory one regards the global symmetry as $U(N)\times U(N)$ with the two diagonal $U(1)$ factors acting trivially on the theory and one adds a level $N$ $\cN=4$   background mixed Chern--Simons term, or $\cN=4$ BF term at level $N$ (see \cite{Kapustin:1999ha}), for the two corresponding $U(1)$ background vector multiplets. This does not modify the 3d theory but it becomes important when we gauge the $U(N)$ global symmetry as we explain now.\footnote{In most of the literature on the topic, the theory is referred to as $T[SU(N)]$ and the presence or not of such background Chern--Simons terms is irrelevant.}

\begin{figure}[h!]
\centering
\includegraphics[scale=0.8]{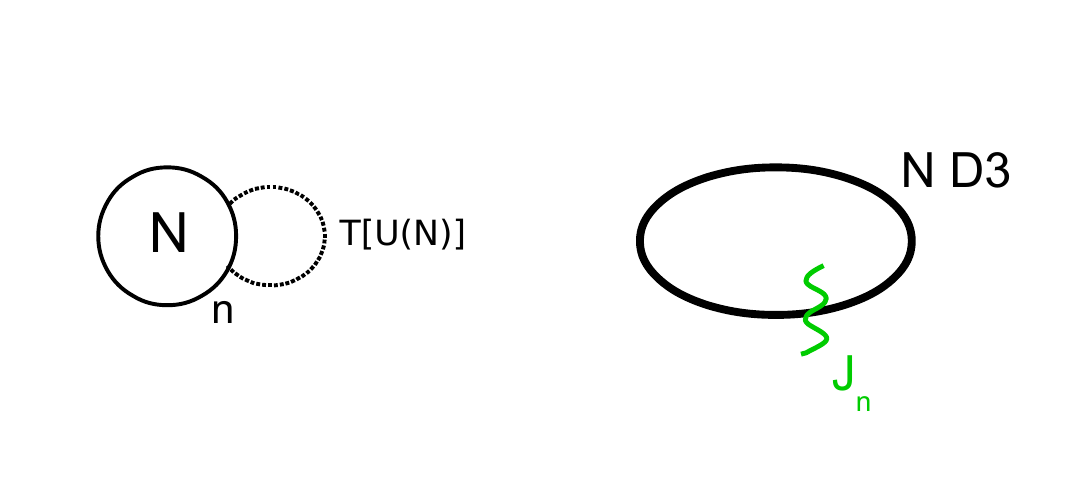} 
\vskip -1cm
\caption{\footnotesize Quiver description of the $J_n$ theories and their brane realization. The subscript $n$ is the Chern--Simons level of the $U(N)$ node.}
\label{JnQuiver}
\end{figure}

This 3d interface theory -- the $T[U(N)]$ theory plus a level $n$ CS term for one flavor $U(N)$ --  is then coupled to the 4d ``bulk" theory by gauging one $U(N)$ global symmetry with the 4d $U(N)$ bulk vector multiplet living on one side of the 3d defect and the other $U(N)$ global symmetry with the 4d $U(N)$ bulk vector multiplet living on the other side. The reason why this description is not fully Lagrangian is that the two $U(N)$ global symmetries are not both present in the UV Lagrangian description of $T[U(N)]$.
As we flow to the infrared the theory becomes three-dimensional and the ``bulk" vector multiplets on both sides of the 3d interface get identified. The resulting 3d theory is shown in Figure \ref{JnQuiver}. It has a single $U(N)$ gauge node with a supersymmetric Chern--Simons coupling at level $n$ and a ``self-coupling" to the $T[U(N)]$ theory.\footnote{The $J_n$ theories (and other $J$-theories) were already considered in the context of the 3d-3d correspondence in \cite{Terashima:2011qi} (section 4.1) -- see also section 5.2 of \cite{Gang:2015wya}-- where they were realized by twisted compactification of the 6d (2,0) theory on a torus bundle over $S^1$. There are still small differences: the gauge nodes are $SU(N)$ instead of $U(N)$ and an $\cN=2$ adjoint mass term is turned on (punctured torus bundle).}

There is a subtlety about the gauging procedure of the $T[U(N)]$ global symmetries that deserves a comment. We are identifying (and gauging) the two $U(N)$ global symmetries of $T[U(N)]$, however there are two possibilities for doing so, namely breaking the symmetry to $U(N)_+ =$ diag$(U(N)\times U(N))$ or $U(N)_- =$ diag$(U(N)\times U(N)^\dagger)$. A natural convention is to associate these two choices to duality interfaces labeled by $S$ and $-S$ respectively. Since $J_n= (-S)T^n$ we will think of the $J_n$ theory as defined with $U(N)_-$ gauging and with CS level $n >2$. In the previous section we mentioned the existence of a $\bar J_n$ solution with $\bar J_n = - J_{-n} = S T^{-n}$. The CFT dual of that solution would have a $T[U(N)]$ theory with $U(N)_+$ gauging and CS level $-n < -2$.
This discussion will have a counterpart in localization computations in the next section.\footnote{Such a distinction between $U(N)$ gauging procedures was already discussed in \cite{Assel:2014awa}.}

The UV description of the $J_n$ theory has a Chern--Simons term at level $n$. Chern--Simons terms naively break the supersymmetry to $\N=3$, however in certain circumstances the supersymmetry enhances to $\N=4$ or more in the infrared limit \cite{Gaiotto:2008sd,Hosomichi:2008jd}. Since we were able to construct the $J_n$ theory in an $\N=4$ preserving fashion, as a compactification of a half-BPS Janus theory, we know that the infrared SCFT has indeed $\N=4$ supersymmetry. This is confirmed by the gravity dual solution, which has this amount of supersymmetry as well. 

The $SU(2)_R$ R-symmetry of the 3d $\N=3$ UV theory must be enhanced at low energies to the $SU(2)\times SU(2)$ R-symmetry of an $\N=4$ SCFT, represented in the gravity dual solution as the isometries of the two 2-spheres. 
 It is interesting to notice that the $J_n$ SCFTs have no continuous global symmetries besides the $\cN=4$ R-symmetry. One may regard them as minimal SCFTs with $\N=4$ supersymmetry in this respect. Correspondingly the supergravity dual backgrounds are very simple, in the sense that do not have five-brane sources.

The 3d SCFTs and their gravity duals are supposed to be two low-energy descriptions of a very simple brane configuration in type IIB string theory, where we have a stack of $N$ D3-branes wrapping a circle with $J_n$ duality twist, as shown in Figure \ref{JnQuiver}.

\subsection{Test of holography}
\label{ssec:TestJanus}

To test the holographic correspondence we compare the on-shell action of the $J_n$-fold Janus solution to the free energy of the $J_n$ theories in the limit where the gravity approximation is valid, which turns out to be the usual large $N$ limit.

The regularized on-shell action was evaluated in \cite{Assel:2012cp,Assel:2012cj} in terms of the harmonic functions $h_1,h_2$, using a consistent truncation to pure gravity. It is given by the remarkably simple formula (in the convention $\alpha'=4$)
\bea
S_{IIB} &=  - \frac{1}{(2\pi)^3} \int_{\Sigma} dx dy \, h_1h_2 \p_z\p_{\bar z} (h_1 h_2) \cr
& =  - \frac{1}{(2\pi)^3} \int_{0}^T dx \int_0^{\frac{\pi}{2}} dy \, h_1h_2 \p_z\p_{\bar z} (h_1 h_2)\,.
\label{OnShellAction}
\eea
The $SL(2,\bR)$ transformation $M$ used to define the $J_n$ solution does not change the on-shell action, therefore we can directly use the above formula with the $h_1,h_2$ functions of the extremal Janus solution \eqref{JanusExtremal}. We obtain
\be
S_{IIB} = \frac{L^8 T}{2^{13} \pi^2} = \frac 12 N^2 \ln\lp \frac 12 \lp n + \sqrt{n^2 - 4} \rp \rp \,.
\label{SIIBJn}
\ee
We would like to know in which regime this result can be compared with the field theory free energy. The type IIB action does not receive quantum or string corrections at the two derivative order.
For the higher derivative corrections to the IIB action to be suppressed we require that $R_{(s)}$ and $g_{(s)}^{\mu\nu}\nabla_\mu \phi\nabla_\nu \phi$ be small, where the index $(s)$ indicates that we use the string frame metric, $g_{(s)\mu\nu} = g_{\mu\nu} e^{\phi}$. We have the relation $g_{(s)}^{\mu\nu}\nabla_\mu \phi\nabla_\nu \phi = g^{\mu\nu}\nabla_\mu e^{-\frac{\phi}{2}}\nabla_\nu e^{-\frac{\phi}{2}}$.

The idea behind these conditions is the following. The string theory action has various terms $S_{k,i}$ with $k>2$ derivatives. Each of these consists of a combination of curvature and derivatives of $\phi$, with a function of the string coupling $f_{k,i}(e^\phi)$ in front, which receive both perturbative and non-perturbative contributions. If $R_{(s)}$ and $g_{(s)}^{\mu\nu}\nabla_\mu \phi\nabla_\nu \phi$ are smaller than $\epsilon$, we expect $S_{k,i}< f_{k,i} \epsilon^k$. Almost all of the $f_{k,i}$ are unknown, but unless their convergence radius gets smaller and smaller with increasing $k$, there will be an $\epsilon$ small enough that $f_{k,i} \epsilon^k$ will be small for all $k$. Flux terms work in the same fashion.

The metric of the $J_n$ solution scales as $g_{\mu\nu}\sim L^2  \sim \sqrt{N}$ and the inverse dilaton $e^{-2\phi} = {\rm Im}(\tau')$ is independent of $N$. We find that both higher derivative terms are bounded $g_{(s)}^{\mu\nu}\nabla_\mu \phi\nabla_\nu \phi , \, R_{(s)} \lesssim  C \big(\frac{\sinh T}{N}\big)^{1/2}$, with $C$ a positive constant. 
Thus both are small in the limit of large $N$ and finite $T$, and the IIB supergravity approximation should be valid in this regime.

\medskip

The result \eqref{SIIBJn} should be compared with the large $N$ free energy $F=-\ln |Z|$, with $Z$ the three-sphere partition function of the $J_n$ theory. The sphere partition function $Z$ can be computed exactly by supersymmetric localization \cite{Kapustin:2009kz,Hama:2011ea} and the final result is expressed as a matrix model whose integrand is a product of contributions from different ingredients of the theory. We briefly review the results of the localization computation in Appendix \ref{app:ZS3}. We also explain there how to account for the coupling to the $T[U(N)]$ theory in the matrix model. For the $J_n$ theory the matrix model is\footnote{Here we ignore the overall phase of $Z$ which does not play a role in our computation.}
\bea
Z &= \frac{1}{N!} \int d^N\sigma \, Z_{\rm CS}(\sigma) \, Z_{\rm vec}(\sigma) \, Z_{T[U(N)]}(\sigma,-\sigma) \cr
& = \frac{1}{N!} \int d^N\sigma \, e^{i\pi n \sum_{i=1}^N \sigma_i^2} \, \prod_{i<j=1}^N \sh^2(\sigma_{ij}) \, \frac{\sum_{\tau\in S_N} (-1)^\tau  e^{-2\pi i \sum_{i=1}^N \sigma_i\sigma_{\tau(i)}} }{\prod_{i<j=1}^N \sh^2(\sigma_{ij}) }  \cr
& = \frac{1}{N!}\sum_{\tau\in S_N} (-1)^\tau \int d^N\sigma \, e^{i\pi n \sum_{i=1}^N \sigma_i^2} \, e^{-2\pi i \sum_{i=1}^N \sigma_i\sigma_{\tau(i)}} \,.
\label{ZJnMM}
\eea
Remarkably the matrix model becomes very simple (gaussian in fact).
In appendix \ref{app:JnPartFunc} we use matrix model techniques to evaluate this matrix model.\footnote{Note added in version 3: This computation was also performed in \cite[App.~I]{Gang:2015wya}. It was then compared to the large $N$ gravity action of M-theory dual solutions which arise from twisted compactification of M5 branes on three-manifolds \cite{Gauntlett:2000ng,Gang:2014qla,Gang:2014ema}. Although in principle we expect the M-theory solutions to be related to the IIB $J_n$-solutions, such a relation, if it exists, is not obvious. M-theory/IIB duality requires shrinking an isometry direction in IIB, and there is no natural candidate here.} It is sufficiently rare to be emphasized that we are able to evaluate exactly, at finite $N$, the sphere partition function $Z$. Miraculously the parameter $T$ of \eqref{JnParam} pops up in the computation and the final result, up to a phase, is \eqref{ZJnFinalApp}
\be
Z =  \frac{e^{\frac{NT}{2}}}{\prod_{j=1}^N \left( e^{j T} - 1 \right) } = \frac{e^{\frac{-N^2 T}{2}}}{\prod_{j=1}^N \left(1 - e^{-j T} \right) }  \,.
\label{ZJnFinal}
\ee
The free energy is then
\be
F = \frac{N^2}{2} \,  T + \sum_{j=1}^N \ln \left( 1 -  e^{-j T}\right) = \frac 12 N^2 T + O(N^0, e^{-T}) \,.
\ee
The leading order term matches the supergravity on-shell action \eqref{SIIBJn} in the supergravity limit that we found above, i.e. large $N$ and finite $T$, providing a very non-trivial test of the holographic duality that we proposed. We observe that the two results also match in the limit of large $T$ and finite $N$. In the CFT dual theory, it corresponds to the limit of large Chern--Simons level $n$ and finite $N$. This suggests that the limit is also a long-wavelength approximation, although this does not follow from our simple analysis. A more complete treatment of the supergravity higher derivative corrections would be needed to explain this observation.

\subsection{Other $J$-fold theories}
\label{ssec:GeneralJanusfold}

We can find other solutions $(J,M(J),T(J))$ to \eqref{SfoldCondition}, allowing for the definition of new compactifications of the extremal Janus solution with twisting by other $J\in SL(2,\bZ)$. 
Taking the trace of \eqref{SfoldCondition} yields the relation
\be
\tr J = e^{T_J} + e^{-T_J} \,,
\label{traceJ}
\ee
(with $T_J \equiv T(J)$) which implies the constraint
\be
\tr J > 2 \,.
\label{Jcondition}
\ee
This excludes for instance $S$ and $T^k$ as duality elements to perform the quotient. Elements satisfying $|\tr J| > 2$ are called {\it hyperbolic}, therefore the condition \eqref{Jcondition} restricts to hyperbolic elements with positive trace.

We can try to solve for the matrix $M$ in \eqref{SfoldCondition} for a given $J$. We find that the condition \eqref{Jcondition} is enough to  find a solution $M(J)$. This means that there is an S-fold solution for all $J\in \SL(2,\bZ)$ satisfying \eqref{Jcondition} and the period $T_J$ is given by the relation \eqref{traceJ}. Explicitly, with $J = \lp \begin{array}{cc} j_1 & j_2 \cr j_3 & j_4 \end{array}\rp $ and $j_1 + j_4 = e^{T_J} + e^{-T_J}$,
\be
M(J) = \lp
\begin{array}{cc}
\frac{\lambda}{1 + e^{-T_J}} & \frac{ j_2}{\lambda (1 - e^{T_J})} \cr
\frac{\lambda j_3}{(1 + e^{T_J})(1 - j_4 e^{-T_J})}  & \frac{(1 - j_4 e^{-T_J})}{\lambda (1 - e^{-T_J})}   
\end{array}\rp \,,
\ee
for any $\lambda \in \bR^\ast$.

Note that (infinitely) many $J$ elements have the same trace and therefore the same period $T$. They are related by $SL(2,\bR)$ transformations (this follows from the relation \eqref{SfoldCondition}), but are not dual in the full string theory, unless the transformation is in $SL(2,\bZ)$.
\smallskip

Once again the 3d dual SCFT can be engineered as the low-energy of a 4d $U(N)$ Janus configuration with pseudo-periodicity $J.\tau(x'+T_J) = \tau(x')$, compactified on a circle with $J$-twisted boundary conditions. The profile of the complexified Yang--Mills coupling $\tau(x')$ is the same as in the supergravity dual solution,
\be
\tau(x') = \frac{\frac{(1 - j_4 e^{-T_J})}{(1 - e^{-T_J})} i \, \lambda' \, e^{2x'} + \frac{ j_3}{(1 + e^{T_J})(1 - j_4 e^{-T_J})}}{\frac{ j_2}{(1 - e^{T_J})}  i \, \lambda' \, e^{2x'} +  \frac{1}{1 + e^{-T_J}}  }\,,
\ee
$\lambda'\in \bR^\ast$, and is of the form \eqref{JanusProfiles} with $a=\frac 12 \left( \frac{j_3}{e^{T_J}-j_4} - \frac{e^{T_J}-j_4}{j_2} \right) $, $D= \frac 12 \left( \frac{j_3}{e^{T_J}-j_4} + \frac{e^{T_J}-j_4}{j_2} \right)$ and $e^{i\psi(x)}= \frac{1 - e^{T} - i \lambda' j_2 e^{2x'}(1 + e^{-T_J})}{1 - e^{T} + i \lambda' j_2 e^{2x'}(1 + e^{-T_J})}$, preserving 3d $\N=4$ supersymmetry.
\smallskip

The 3d $\N=4$ SCFT dual theories can be described in a more practical way as the infrared limit of a 3d quiver using the approach of \cite{Gaiotto:2008ak}. First we need to express the duality element $J$ as a product\footnote{This is the most general form of an $SL(2,\bZ)$ element up to conjugation by $T^k$ and $S$.}$^{,}$\footnote{Despite the conflicting notation, the $SL(2,\bZ)$ matrix $T$, should not be confused with the period $T$ appearing in the $J$-fold solution.}
\be
J = \pm (-ST^{n_1})(-ST^{n_2}) \cdots (-ST^{n_p}) = \pm J_{n_1} J_{n_2} \cdots J_{n_p} \,,
\label{Jexpand}
\ee
where $n_i$ are positive integers. The overall sign $\pm$ should be fixed by the requirement $\tr J > 2$.\footnote{The constraint on the trace also restricts the possible values of $n_1,n_2, \cdots ,n_p$. It is also possible to take $n_a \ge 2$, i.e. exclude $n_a=1$, since $TSTST=-S$.}

The 3d CFT is then the infrared limit of a quiver-like circular theory with $p$ $U(N)$ nodes with Chern--Simons terms at levels $n_i$, coupled together via $T[U(N)]$ gaugings. To be precise when coupling a $T[U(N)]$ theory to two $U(N)$ gauge nodes we can identify the $U(N)\times U(N)$ nodes with the global symmetries $U(N)\times U(N)$ or $U(N)\times U(N)^{\dagger}$ of $T[U(N)]$ (i.e. $S$ or $-S$ interfaces), leading to many choices. However due to the freedom in redefining what we mean by $U(N)$ and $U(N)^\dagger$, there are only two globally inequivalent choices corresponding the choice of $\pm$ in \eqref{Jexpand}. If the sign is $+$ we pick all gaugings with $U(N)\times U(N)^{\dagger}$ and if the sign is $-$ we pick one gauging with $U(N)\times U(N)$ and the others with $U(N)\times U(N)^{\dagger}$.\footnote{Note that with the $U(N)\times U(N)^{\dagger}$ gauging, the level $N$ BF term which is part of the definition of the $T[U(N)]$ theory becomes a level $-N$ BF term for the diagonal $U(1)$s of the two gauge nodes connected by the $T[U(N)]$ link.} 
In the abelian case $N=1$, these theories reduce to the Chern--Simons quivers that were studied in \cite{Ganor:2014pha}.

\begin{figure}[h!]
\centering
\includegraphics[scale=0.8]{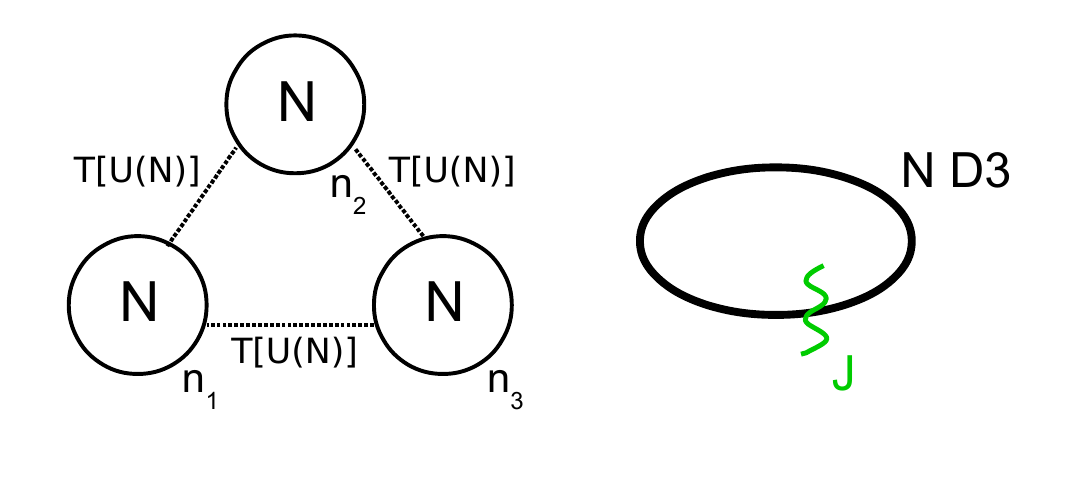} 
\vskip -0.5cm
\caption{\footnotesize Quiver description of a $J$ theory, with $J=J_{n_1}J_{n_2}J_{n_3}$, and its brane realization. Subscripts of gauge nodes indicate Chern--Simons levels.}
\label{JQuiver}
\end{figure}

An example with three nodes is shown in Figure \ref{JQuiver}.

Here again the SCFT has naively only $\N=3$ supersymmetry due to the presence of the Chern--Simons terms, however the supersymmetry must be enhanced to $\N=4$ in the infrared limit, since we constructed it from a compactification of a 4d Janus configuration preserving $\N=4$ supersymmetry, and the gravity dual solution has indeed the corresponding 16 Killing spinors. 

Our construction does not provide holographic dual solutions for $J$-fold theories with $J$ elliptic ($|\tr J| < 2$) or parabolic ($|\tr J|=2$). For the parabolic case $J=T^k$, the 3d theory is simply $\N=3$ $U(N)$ Chern--Simons theory, which after integrating out auxiliary fields is a pure Chern--Simons theory, with no local degrees of freedom.
For elliptic elements there are no known gravity duals.\footnote{A construction of 4d SYM Janus configurations on a circle with topological twist involving elliptic elements was presented in \cite{Ganor:2008hd,Ganor:2010md,Ganor:2012mu}.} It could be that these theories do not flow to SCFTs.

\medskip

\noindent{\bf Holographic test}:

The evaluation of the on-shell supergravity action is identical to that of the $J_n$ theory with the period $T:= T_J > 0$ defined by $\tr J := e^{T_J} + e^{-T_J}$,
\be
S_{IIB} = \frac 12 N^2 T_J + O(N^0) \,.
\label{SIIBJ}
\ee
On the CFT side, the sphere partition function $Z_J(N)$, for $J = J _{n_1}\cdots J_{n_p}$, is computed by the matrix model\footnote{This form of the matrix model arises after the cancellation between the vector multiplet factors \eqref{VecFactor} and the denominators of the $T[U(N)]$ factors \eqref{TUNFactor}. Moreover the $p$ sums over $S_N$ permutations arising from the $p$ $Z_{T[U(N)]}$ factors simplify to a single sum by redefinitions of the eigenvalues.}
\be
Z_J = \frac{1}{N!}\sum_{\tau\in S_N} (-1)^\tau \int \left(\prod_{a=1}^p d^N\sigma_a \, e^{i\pi n_a \sum_i \sigma_{a,i}^2}\right) \left(\prod_{a=1}^{p-1} e^{-2\pi i \sum_i \sigma_{a,i}\sigma_{a+1,i}} \right) e^{-2\pi i \sum_i \sigma_{p,i}\sigma_{1,\tau(i)}} \,.
\label{ZJMM}
\ee
For $J=  -J_{n_1}\cdots J_{n_p}$, the last factor in the integrand becomes its complex conjugate $e^{2\pi i \sum_i \sigma_{p,i}\sigma_{1,\tau(i)}}$, accounting for the different $U(N)\times U(N)$ gauging of one $T[U(N)]$ factor, as explained above. Once again we discard a possible phase factor of the matrix model.

Let us consider $J=J_{[2]}=J_{n_1}J_{n_2}$. The condition on the trace is $\tr J_{[2]} = n_1n_2 -2 > 2$. The partition function is
\be
Z_{[2]} = \frac{1}{N!}\sum_{\tau\in S_N} (-1)^\tau \int d^N\sigma_{1} d^N\sigma_2 \, e^{i\pi n_1 \sum_{i=1}^N \sigma_{1,i}^2} e^{i\pi n_2 \sum_{i=1}^N \sigma_{2,i}^2}\, e^{-2\pi i \sum_{i=1}^N (\sigma_{1,i} + \sigma_{1,\tau(i)})\sigma_{2,i}} \,.
\ee
Integrating out $\sigma_{2,i}$ and rescaling $\sigma_{1,i} = \sqrt{n_2} \, \sigma_i$, we obtain
\be
Z_{[2]} = \frac{e^{\frac{i\pi N}{4}}}{N!}\sum_{\tau\in S_N} (-1)^\tau \int d^N\sigma  \, e^{i\pi (n_1n_2-2) \sum_{i=1}^N \sigma_{i}^2}  \, e^{-2\pi i \sum_{i=1}^N \sigma_i\sigma_{\tau(i)}} \,,
\ee
matching the partition function of the  $J_{n_1n_2-2}$-fold theory (up to a possible phase). The free energy is thus
\be
F_{[2]} = - \ln Z_{[2]} = -\ln  Z[J_{n_1n_2-2}] = \frac{1}{2}N^2 T_{J_{n_1n_2-2}} + O(N^0) \, = \, \frac{1}{2}N^2 T_{J_{[2]}} + O(N^0)  \,,
\ee
where we used $T_{J_{n_1n_2-2}}= T_{J_{[2]}}$, due to $\tr J_{[2]} = \tr J_{n_1n_2-2} =  n_1n_2-2$. We find agreement with the on-shell action evaluation \eqref{SIIBJ}.

We can compute the general $Z_{[p]} \equiv Z_J$ for $J=J_{n_1} J_{n_2} \cdots J_{n_p}$ as follows. Introduce the matrix 
\begin{equation}\label{eq:Qmat}
	Q= \left(\begin{array}{cccccc}
		 n_1 & -1  & 0 & \cdots & 0 & -1\\ 
		 -1 & n_2 & -1 & \cdots & & 0 \\
		0 & -1 & n_3 & \ddots & & \vdots \\
		\vdots & & \ddots & \ddots & \ddots & 0\\
		0 & & & \ddots &\ddots &-1\\
		-1 & \cdots & & 0 & -1 & n_p
 	\end{array}\right)\,,
\end{equation}
and call $M\equiv \mathrm{Min}_{11}Q \equiv (M^{ab})_{a=2,\cdots,p \atop b=2,\cdots,p}$, where we use $\mathrm{Min}_{ab}$ to denote the matrix obtained by deleting the $a$-th row and $b$-th column, without additional signs. Moreover we will call $\mathrm{min}_{ab} = \det \mathrm{Min}_{ab}$. Now we can write
\begin{equation}
	Z_{[p]}= \sum_{\tau\in S_N} \frac{(-1)^\tau}{N!} \int d^N \sigma_1 \ldots d^N \sigma_p \, e^{i\pi (n_1 \sigma_1^2 +  M^{ab} \sigma_a \cdot \sigma_b -2 \sigma_a \cdot (\delta^a{}_{2}\sigma_1 + \delta^a{}_{p} \sigma^\tau_1))}\,,
\end{equation}
with the notations $\sigma_a^2 = \sum_{i=1}^N \sigma^2_{a\, i}$ , $\sigma_a.\sigma_b = \sum_{i=1}^N \sigma_{a\, i} \sigma_{b\, i}$ and $\sigma^{\tau}_i = \sigma_{\tau(i)}$.
By performing the integral over $\sigma_2,\ldots,\sigma_p$, discarding a phase $e^{\frac{i\pi (p-1)N}{4}}$, and rescaling $\sigma_1$ with $\sigma_1 = (\det M)^{1/2}\sigma$ we obtain
\begin{align}
	Z_{[p]}&= \sum_{\tau\in S_N} \frac{(-1)^\tau}{N!} \int \frac{d^N \sigma_1}{(\det M)^{N/2}} \, e^{i\pi(n_1 \sigma_1^2-(\delta^a{}_{2}\sigma_1 + \delta^a{}_{p} \sigma^\tau_1)M^{-1}_{ab}(\delta^b{}_{2}\sigma_1 + \delta^b{}_{p} \sigma^\tau_1))}\nonumber\\
	&= \sum_{\tau\in S_N} \frac{(-1)^\tau}{N!} \int d^N \sigma \, e^{i\pi( \det M(n_1 - M^{-1}_{pp}-M^{-1}_{22}) \sigma^2 -2(\det M )M^{-1}_{2p} \sigma \cdot \sigma^\tau)}\nonumber\\
	&= \sum_{\tau\in S_N} \frac{(-1)^\tau}{N!} \int d^N \sigma \, e^{i\pi( ((\det M) n_1 - \mathrm{min}_{pp}M-\mathrm{min}_{22}M) \sigma^2 -2 (-1)^k\mathrm{min}_{2p} M \sigma \cdot \sigma^\tau)} \nonumber\\
	&=\frac{1}{N!}\sum_{\tau\in S_N} (-1)^\tau\int d^N \sigma \, e^{i\pi( (\det Q+2) \sigma^2 -2\sigma \cdot \sigma^\tau)}\,. \label{eq:Zksteps}
\end{align}
The identities we used in the last steps are straightforward (if a little involved) applications of the usual formulas for determinants. The last expression can be recognized as the partition function of the $J_q$ theory with $q=\det Q + 2$,  which we evaluated in appendix \ref{app:JnPartFunc},
\be
Z_{[p]} = Z[J_{q}] = \frac{1}{2}N^2 T_{J_{q}} + O(N^0) \,,
\label{eq:Zkeval1}
\ee
with $T_{J_{q}}$ defined by the relation $\tr J_{q} \equiv q = e^{T_{J_{q}}} + e^{-T_{J_{q}}}$.
To complete the computation we need to show  $T_{J_{q}} = T_J$, or equivalently $\tr J_q = \tr J$. Explicitly we need to show
\begin{equation}\label{eq:detQ}
	\det Q + 2 = \mathrm{Tr} J_{n_1}\ldots J_{n_p} \,.
\end{equation}
Both the left- and the right-hand-side of this identity are manifestly cyclic polynomials in $n_1,\ldots,n_p$: they have the property that $P(n_1,n_2,\ldots,n_p)=P(n_2,\ldots,n_p,n_1)$. Such a polynomial is uniquely determined by its restriction to equal values $P(\ti n,\ldots, \ti n)$. Thus we only have to prove (\ref{eq:detQ}) when $n_1=\ldots=n_p= \ti n$. In fact in this case $Q$ is equal to $C_{p,\ti n}$, the matrix we used in (\ref{eq:Cell}) to compute the partition function in the $p=1$ case. (In that context the matrix was acting on a variable $\sigma$ with $\ell$ components). So we can use (\ref{eq:detcl}), using again the trick of writing $\ti n= e^{\ti T}+e^{-\ti T}$. We obtain $2+\det Q|_{n_1=\ldots=n_p=\ti n}= e^{p\ti T}+ e^{-p\ti T}$. On the other hand, $J_{\ti n}$ has eigenvalues $e^{\pm \ti T}$, so $\mathrm{Tr}(J_{\ti n})^p=e^{p \ti T}+ e^{-p\ti T}$. We have thus proven (\ref{eq:detQ}). Going back to (\ref{eq:Zkeval1}) we obtain
\begin{equation}
	Z_J \equiv Z_{[p]} = \frac{1}{2}N^2 T_{J} + O(N^0) \,
\end{equation}
with $J=J_{n_1}\ldots J_{n_p}$. A similar computation holds for the choice $J = -J_{n_1}\ldots J_{n_p}$.

We obtain a beautiful agreement with the supergravity on-shell action \eqref{SIIBJ} at large $N$.

We notice a posteriori that the appearance of the matrix $Q$ \eqref{eq:Qmat} in the computation should not be a surprise. Indeed this is the matrix of abelian Chern--Simons terms in the $J$-fold theory, up to a multiplicative factor $N$: $n_a N$ are the CS levels of the diagonal $U(1)\subset U(N)$ at each node and $-N$ is the level of the mixed CS term between adjacent gauge nodes, due to the $T[U(N)]$ links. The matrix $Q$ encodes the data of the quiver theory with $T[U(N)]$ links in an efficient way.

\section{S-flip quiver SCFTs}
\label{sec:SfoldQuivers}

We now turn to the construction of a different class of S-fold solutions, which we call S-flips. They are obtained as the quotient of $\N=4$ solutions dual to circular quiver theories by the $S$ element of $SL(2,\bZ)$. The corresponding 3d dual SCFTs will have reduced $\N=3$ supersymmetry and will be circular quiver theories with a $T[U(N)]$ `link'. To construct holographic dual pairs it will be useful to consider the brane realization of 3d $\N=4$ quiver gauge theories, which involves D3-branes wrapping a circle and crossing NS5 and D5-branes \cite{Hanany:1996ie}. Explicitly we will take the D3s along $x^{0123}$ with $x^3$ compact, the NS5s along $x^{012456}$ and the D5s along $x^{012789}$. Such brane setups preserve eight supercharges and the low energy theory on the D3-branes has 3d $\N=4$ supersymmetry. The extra ingredient needed to construct the S-flip theories is an S-interface associated with the $T[U(N)]$ link, and will be responsible for breaking the supersymmetry to $\N=3$.

\subsection{Half-ABJM theory}
\label{ssec:HalfABJM}

\begin{figure}[h!]
\centering
\includegraphics[scale=0.8]{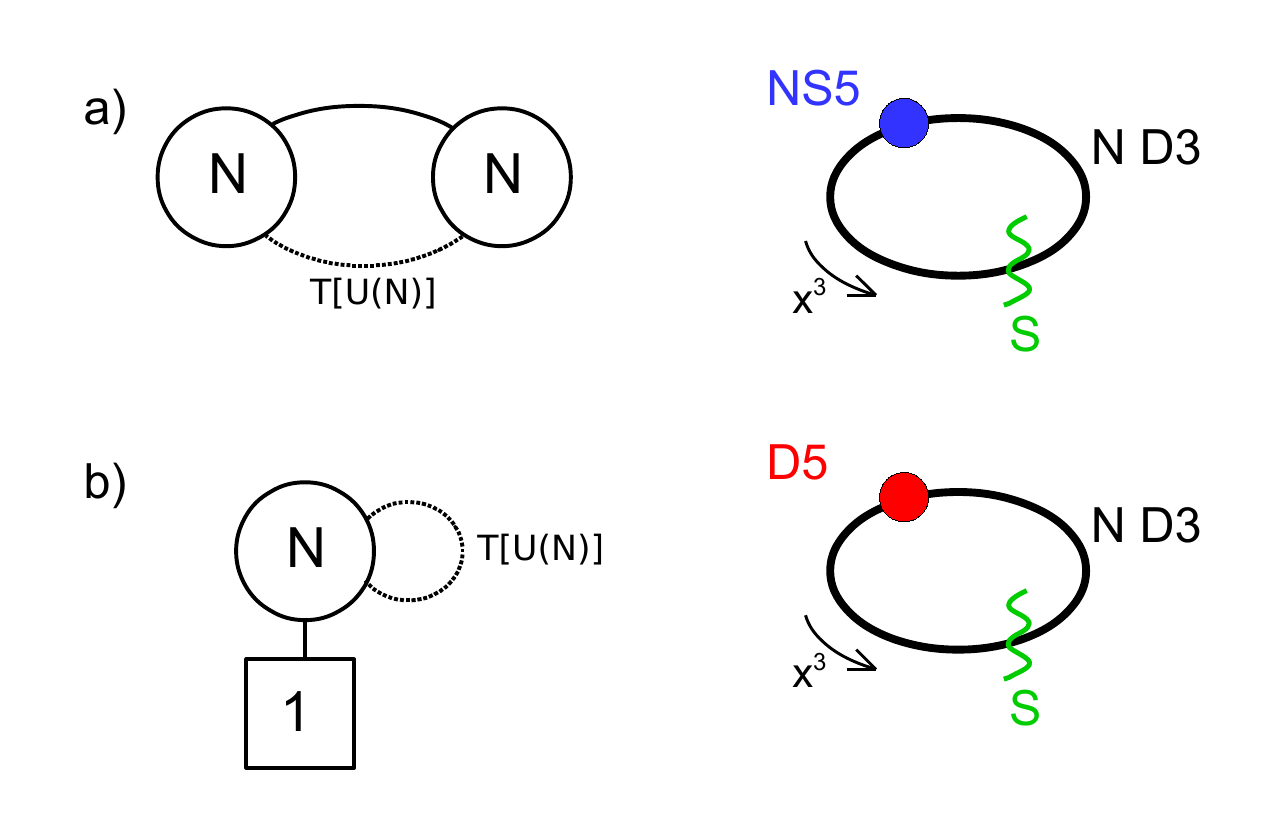} 
\vskip -0.5cm
\caption{\footnotesize a) Half-ABJM quiver and its brane realization. b) Half-ABJM mirror and its brane realization.}
\label{halfABJM}
\end{figure}

The simplest example of such an S-flip 3d quiver theory is the circular quiver with gauge group $U(N)\times U(N)$, with a bifundamental hypermultiplet and with a $T[U(N)]$ link between the two nodes, namely a $T[U(N)]$ theory (as described in Section \ref{ssec:JnfoldSCFTs}) with its $U(N)\times U(N)$ global symmetries gauged by the two $U(N)$ gauge nodes.\footnote{As explained in Section \ref{ssec:JnfoldSCFTs}, there are two inequivalent gaugings of the $T[U(N)]$ global symmetries, leading to two different theories. Here, however, the two theories are related by a parity transformation (which reverses the sign of CS and FI terms in the $T[U(N)]$ theory) and are therefore equivalent. (A parity transformation maps the $J_n$ theory of Section \ref{sec:Janus} to the $\bar{J_n}=-J_{-n}$ theory)} It is mirror dual\footnote{We use the denomination {\it mirror dual} abusively here, since as we will discuss these theories have only $\cN=3$ supersymmetry and there is no notion of Higgs and Coulomb branch exchange, however this duality is still implemented by S duality in the type IIB brane realization, as for the mirror symmetry duality of 3d $\cN=4$ theories.} to the theory with a single gauge node $U(N)$ with one fundamental hypermultiplet and a $T[U(N)]$ link connecting to the same $U(N)$ node. The two quivers are shown in Figure \ref{halfABJM}. We will call these theories half-ABJM  and half-ABJM mirror, for reasons that will be clear shortly. 

The type IIB brane realizations of the half-ABJM theory and its mirror are also shown in Figure \ref{halfABJM}. They involve $N$ D3-branes wrapping the $x^3$ circle and intersecting an NS5-brane and an $S$ interface, or a D5-brane and an S-interface, respectively. 
The S-interface is a monodromy wall in 10d across which the theory undergoes an S-duality action and spacetime is glued with the rotation $R_{S}: (x^{456},x^{789}) \to (x^{789},-x^{456})$ (an order four involution).
The action of S-duality (or mirror symmetry) can be implemented by letting the S-interface wind once around the $x^3$ circle, changing the type of five-brane from NS5 to D5 or vice-versa. Equivalently we can act on the whole brane configuration with a global S-duality action and  the reflection $R_S$, exchanging NS5 and D5, and leaving the S-interface invariant (since $S^{-1}S S=S$).

\begin{figure}[h!]
\centering
\includegraphics[scale=0.8]{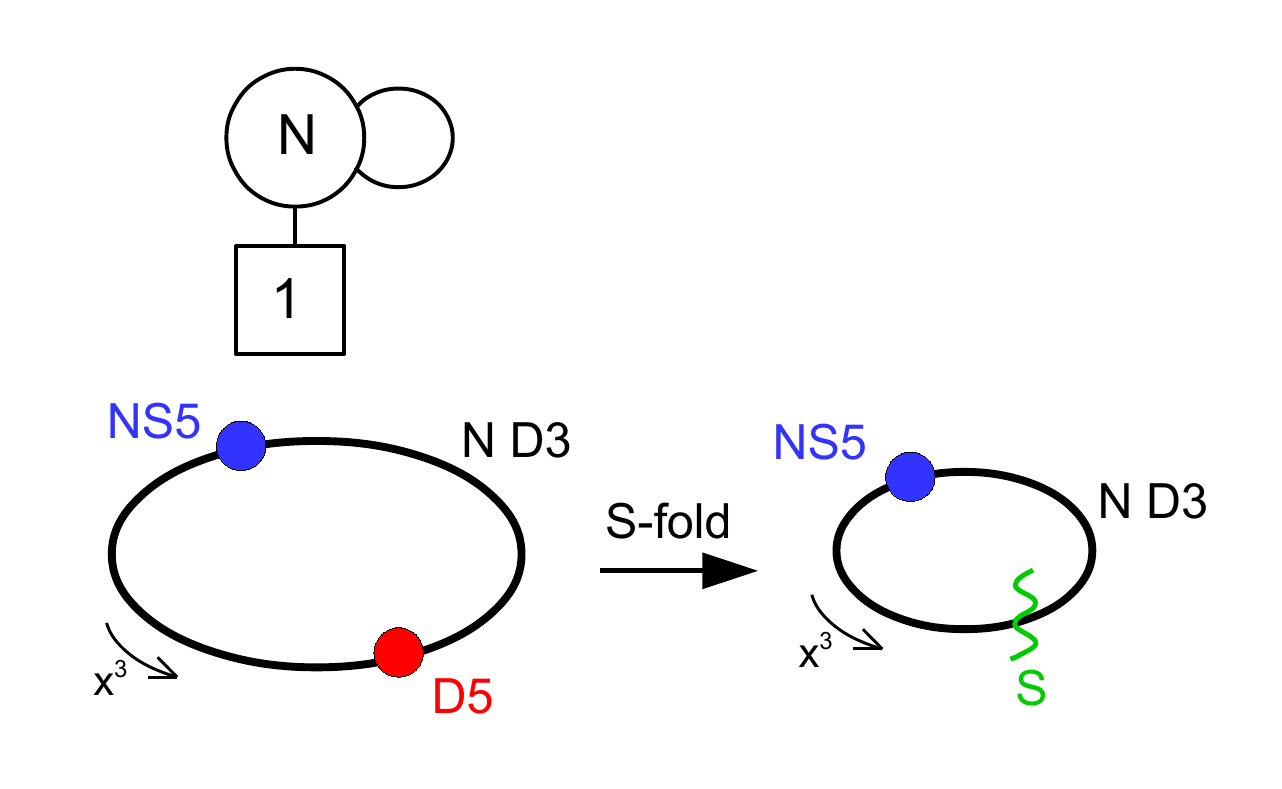} 
\vskip -0.5cm
\caption{\footnotesize A mirror quiver of the ABJM theory at level $k=1$. The S-quotient of its brane realization leads to the brane realization of the half-ABJM theory.}
\label{ABJMSfold}
\end{figure}

A useful point of view is to consider these brane realizations as arising from an S-quotient of a more traditional brane configuration. In this case we can start with the brane configuration with $N$ D3s crossing one NS5 and one D5-branes. This is a type IIB brane realization of the ABJM theory at CS level $k=1$ (or rather a brane realization of a mirror dual theory). This brane setup is invariant under the combined action of $S$ duality and a translation along the $x^3$ circle by a half period. We can then quotient by this action and the resulting brane configuration is that of the half-ABJM theory or its mirror, depending on how we perform the quotient, as shown in Figure \ref{ABJMSfold}.

The infrared SCFT resulting from the S-quotient procedure has only $\N=3$ supersymmetry. This is not immediately obvious from the quiver description of Figure \ref{halfABJM}-a, because 
the quiver theory and the $T[U(N)]$ theory preserve $\N=4$ supersymmetry. The $T[U(N)]$ coupling, described by gauging the two $U(N)$ global symmetries with two $U(N)$ quiver nodes, also involves an $\N=4$ BF coupling \cite{Kapustin:1999ha} between the two diagonal  $U(1)\subset U(N)$ factors. Such a BF coupling preserves $\N=4$ supersymmetry in an unusual way, exchanging the roles of the R symmetry factors $SU(2)_C$ and $SU(2)_H$ for the two $U(1)$ vector multiplets; namely, it preserves $\N=4$ supersymmetry when coupling together a twisted and an untwisted vector multiplet. However, in the present situation the theory is a circular quiver, so that the two $U(1)$ vector multiplets involved in the BF term are also coupled through the rest of the quiver (here through a bifundamental hypermultipet), therefore they are both untwisted vector multiplets from the viewpoint of the circular quiver theory. We thus have a BF term which couples two untwisted vector multiplets and thus preserves only $\N=3$ supersymmetry (as any $\N=3$ Chern--Simons term). This implies that $SU(2)_C$ and $SU(2)_H$ are identified and that the R symmetry is only $SU(2)$. 
This is also visible in the brane construction where the S-interface includes a twist by the space rotation $R_S$. This twist breaks the rotation invariance to $SO(3) =$ diag$(SO(3)_{456}\times SO(3)_{789})$.

 Similar observations about the supergravity dual solution will confirm that the 3d SCFT has $\N=3$ supersymmetry.

\subsection{Supergravity dual background}
\label{ssec:DualBkgrd}

To find the type IIB supergravity solution dual to the half-ABJM theory, we start with that of the (dual of the) ABJM theory at level one, which is part of the class of solutions described in \cite{Assel:2012cj} as the holographic dual backgrounds of 3d $\N=4$ circular quivers. In this class of IIB solutions, the metric is a warped product  ${\rm AdS}_4\times S^2\times S^2\times \Sigma$, with $\Sigma$ a Riemann surface with the topology of an annulus. The solutions are given by the two real harmonic functions $h_1, h_2$ on $\Sigma$,
\bea
\label{h1h2halfABJM0}
h_1(z,\bar z)  &= -  \gamma \ln\left[ \frac{\vartheta_{1}\left(\frac{\pi}{2} + i(z - \delta) \big| \frac{it}{\pi} \right)}{\vartheta_{2}\left( \frac{\pi}{2} + i(z - \delta)  \big| \frac{it}{\pi}  \right)} \right] + c.c. \,,\cr
h_2(z,\bar z)  &= -  \hat \gamma \ln \left[ \frac{\vartheta_{1}\left(-i(z -\hat\delta) \big| \frac{it}{\pi} \right)}{\vartheta_{2}\left(-i(z -\hat\delta) \big| \frac{it}{\pi} \right)} \right] + c.c.  \,,
\eea
where we used the Jacobi Theta functions\footnote{We use the convention $q = e^{\pi i \tau}$.}
\bea
\vartheta_{1}(z | \tau ) &= 2 q^{\frac{1}{4}} \sin (z/2) \prod_{n=1}^\infty (1-q^{2n})(1 - q^{2n} e^{i z})(1 - q^{2n} e^{- i z}) \ , \cr
\vartheta_{2}(z | \tau) &= 2 q^{\frac{1}{4}} \cos (z/2) \prod_{n=1}^\infty (1-q^{2n})(1 + q^{2n} e^{iz})(1 + q^{2n} e^{-iz}) \ .
\eea
The complex coordinate $z=x+iy$ on the annulus has periodicity $z \sim z + 2t$, with $t \in \bR_{>0}$ and range $0 \le y \le \frac{\pi}{2}$ along the vertical axis.
The parameters of the solution are all quantized by the flux quantization conditions. In the case at hand we have a D5 singularity on the upper boundary with one unit of D5 flux and no D3 flux emanating from it, and an NS5 singularity on the lower boundary with one unit of NS5 flux and no D3 flux emanating from it, leading to\footnote{The D3 fluxes in such geometries are subject to ambiguities, related to large gauge transformations of the $B_2$ and $C_2$ form fields. The discussion here implies a certain choice of gauge for these fields (see \cite{Assel:2012cj}).}
\bea
& N_{D5} = 1 = \gamma  \,, \quad \hat N_{NS5} = 1 = \hat\gamma \,, \cr
& N_{D3\to D5} = \hat N_{D3 \to NS5} = 0 \  \to \  \delta - \hat\delta = t \,, 
\eea
using the non-standard convention $\alpha'=4$ (following \cite{Assel:2012cj}).  Common shifts of the $\delta$ and $\hat\delta$ along $x$, translating five-brane stacks, are immaterial, and we fix $\delta = -\hat\delta = \frac t2$.
In addition there are $N$ units of D3 flux wrapping the annulus, with the relation 
\bea
N &= \frac{2}{\pi} \sum_{n\ge 0} (2n+1)\, \text{arctan}\lp e^{-2(n + 1/2)t}\rp  \,,
\label{NtRel}
\eea
which fixes $t$ as a function of $N$.
Thus the solution reads
\bea
\label{h1h2halfABJM}
h_1(z,\bar z)  &= -   \ln\left[ \frac{\vartheta_{1}\left(\frac{\pi - i t}{2} + i z \big| \frac{it}{\pi} \right)}{\vartheta_{2}\left(\frac{\pi - i t}{2} + i z  \big| \frac{it}{\pi}  \right)} \right] + c.c. \,,\cr
h_2 (z,\bar z) &= -  \ln \left[ \frac{\vartheta_{1}\left(-i(z +\frac t2) \big| \frac{it}{\pi} \right)}{\vartheta_{2}\left(-i(z +\frac t2) \big| \frac{it}{\pi} \right)} \right] + c.c.  \,.
\eea

The action of S-duality on such a type IIB solution is implemented essentially by the exchange of the two harmonic functions $S: (h_1,h_2) \to (h_2,h_1)$.
To go back to the convention where D5 singularities are on the upper boundary of $\Sigma$ and NS5 singularities on the lower boundary, one can combine the S action with the symmetry $z\to \frac{i\pi}{2} -z$ on $\Sigma$. The corresponding $S'$-duality action is then
\bea
& S': \quad  h_1(z) \to h_2 \left(\frac{i\pi}{2} -z\right)    \cr
& \phantom{S: \quad} \,  h_2(z) \to h_1 \left(\frac{i\pi}{2} -z\right) \,.
\label{Sprimeaction}
\eea
In the flat brane picture this transformation can be associated with S-duality combined with a reflection $x^3 \to - x^3$ of the circle direction. This is not quite the S action that we are looking for.
There is another way to define an S-duality action for the solutions on the annulus $\Sigma$, which combines the S action with the symmetry $z \to \frac{i\pi}{2} + t + \bar z$, namely a reflection in the $y$ direction and a translation by a half period $t$ along $x$,
\bea
& \cS: \quad  h_1(z) \to h_2 \left(\frac{i\pi}{2} + t + \bar z\right)  \cr
& \phantom{S: \quad} \,  h_2(z) \to h_1 \left(\frac{i\pi}{2} + t + \bar z\right) \,.
\label{Saction}
\eea
This $\cS$ duality action is suitable for our purposes, because it has no fixed points on $\Sigma$ and thus allows for taking the quotient without introducing singularities. Thus this is the notion of S action on the supergravity solution that we retain. It is also naturally identified with the S action on the flat brane configuration that we discussed above, which involves a translation along the circle direction (here identified with the direction $x$).

It is not hard to see that the solution \eqref{h1h2halfABJM} is invariant under $\cS$, as expected from the analysis of the brane configuration. 
This invariance allows us to consider the quotient of the solution by $\cS$, which according to the simple discussion above should be the gravity solution dual to the half-ABJM theory. 

To describe the $\bZ_2$ quotient solution we need to choose a fundamental domain of the $\cS$ action in the surface $\Sigma$. There are various choices, but a simple one is to take the fundamental domain $\Sigma' =\{ z=x+iy \, | \, 0 \le x < t, \, 0\le y\le \frac{\pi}{2}\}$, with the identification $x \sim x + t$ (half the period of the initial solution). The local solution is still given by \eqref{h1h2halfABJM} on the patch $\Sigma'$, and the values on the vertical boundary $x=t \sim 0$ are related by an S-duality transformation $(h_2,h_1)(t,\frac{\pi}{2}-y) = (h_1,h_2)(0,y)$. The global solution is therefore given by the pair $(h_1,h_2)$ being a section of a non-trivial S-bundle ($\bZ_2$ bundle). This is our first construction of an S-flip solution. 

Notice that the S-gluing at $x=t\sim0$ involves a reflection along the $y$ axis $y\to \frac{\pi}{2}-y$ and consequently $\Sigma'$ has the topology of a M\"obius strip. In particular $\Sigma'$ has a single boundary. This will be generic in S-flip solutions, which can be associated to the solutions with internal Riemann surface $\Sigma$ having the topology of a M\"obius strip. 

There is an additional subtlety related to the geometric action on the two $S^2$. In the $\cS$ action we exchange the two harmonic functions $h_1,h_2$. This turns out to correspond to the action of S-duality on the three-form fluxes $H_3,F_3$, which transform as a doublet, combined with the exchange of the two $S^2$.\footnote{More precisely the volume forms $\omega_{i=1,2}$ on the two $S^2$ are transformed as $(\omega_1,\omega_2)\to (-\omega_2,\omega_1)$. Together with the reflection along $y$, this geometric action preserves the orientation in the full geometry. This combined geometric action is the `near horizon limit' of  the rotation $R_S$ in the flat brane picture.} This implies that the action of $\cS$ on the geometry includes  both the transformation $z \to \frac{i\pi}{2} + t + \bar z$ discussed above and the exchange of the two $S^2$. Then, in the S-fold geometry the gluing conditions at $x=t \sim 0$ include the permutation of the two $S^2$. This implies that the $SO(3)\times SO(3)$ rotation symmetry is broken to the diagonal $SO(3)$ in the S-fold solution. The isometries of the solution are identified with the R-symmetry of the dual 3d CFT, therefore the dual half-ABJM SCFT has only $\N=3$ supersymmetry, corresponding to a single $SU(2)$ R-symmetry, paralleling nicely the gauge theory analysis.

\begin{figure}[h!]
\centering
\includegraphics[scale=1]{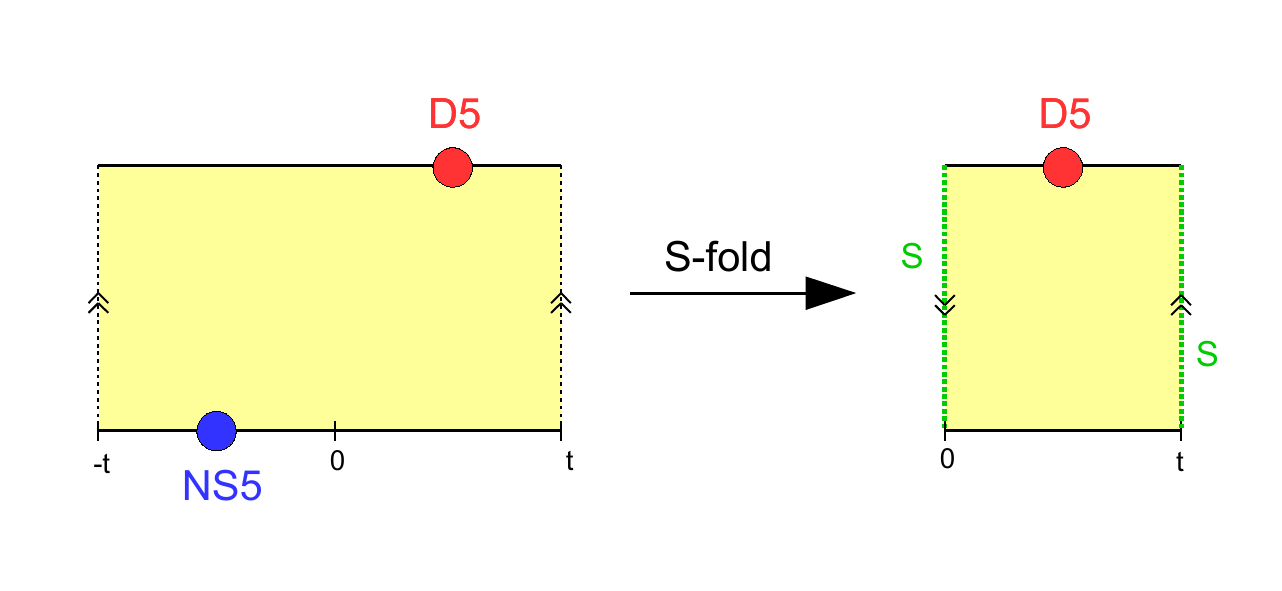} 
\vskip -0.5cm
\caption{\footnotesize On the left: the $\Sigma$ annulus (yellow) with a D5 (red) and an NS5 (blue) singularity, corresponding to the initial ABJM ($k=1$) supergravity solution. On the right: the $\Sigma'$ M\"obius strip with a single D5 singularity and an S-interface (green dashed line), corresponding to the solution quotiented by $\cS$ in \ref{Saction}.}
\label{Sfolding}
\end{figure}

The $\bZ_2$ quotient is described in Figure \ref{Sfolding}. In the resulting S-flip solution, there is a single D5 singularity on the boundary and an S-interface. By moving the S-interface along the $x$ axis by a period $t$, the D5 singularity gets traded for an NS5 singularity, similarly to what happened when moving the S-interface around the $x^3$ circle in the flat brane picture. This can be understood as the action of 3d mirror symmetry in the gauge theory. In the supergravity solution it is a simple change of S-duality frame.

\subsection{Solutions for S-flip quivers}
\label{ssec:GenSol}

The construction of the previous section can be generalized, starting from circular quiver theories whose supergravity solution is invariant under the action $\cS$ \eqref{Saction} and quotienting by $\cS$. 
The 3d $\N=3$ SCFTs dual to these S-fold solutions will comprise the IR limit of all `good' circular quivers with a $T[U(N)]$ link connecting two nodes of minimal rank.

Good quivers refer to quiver theories where the number of fundamental flavors at each node bigger or equal to twice the number of colors, $N_{f,i} \ge 2 N_i$.\footnote{The other quiver theories with unitary gauge nodes (`bad' theories) a priori do not admit new fixed point SCFTs and therefore there is no supergravity solution associated to them (see \cite{Assel:2017jgo,Assel:2018exy} for a recent discussion on the fixed points of bad theories).} They were labeled in \cite{Assel:2012cj} by a triple ($\rho,\hat\rho,N$), where $\rho,\hat\rho$ are two ordered partitions of an integer $\ti N$ satisfying $\rho^T \ge \hat\rho$ and $N>0$.\footnote{Here $\rho$ and $\hat\rho$ are viewed as Young tableaux and the inequality means that the sum of the boxes in the first $n$ rows of $\rho^T$ is bigger than or equal to the sum of the boxes in the first $n$ rows of $\hat\rho$, for all $n$.} The corresponding infrared fixed points were dubbed $C^\rho_{\hat\rho}(SU(\ti N),N)$. In this description $N$ is the lowest rank among the gauge nodes in the quiver and $(\rho,\hat\rho)$ repackage the data of the remaining node ranks and the numbers of fundamental hypermultiplets for each node. We refer to \cite{Assel:2012cj} for a precise dictionary between the gauge theory data and the triple $(\rho,\hat\rho,N)$.
In the supergravity solution associated to $C^\rho_{\hat\rho}(SU(\ti N),N)$,  the partition $\rho$ describes the D3 fluxes emanating from D5-brane stacks on the upper boundary of the annulus $\Sigma$ and $\hat\rho$ describes the D3 fluxes emanating from NS5-brane stacks on the lower boundary of $\Sigma$. The data of the two partitions $(\rho,\hat\rho)$ is encoded in the supergravity solution in two sets of parameters $(\gamma_a,\delta_a)$ and $(\hat\gamma_b,\hat\delta_b)$, associated to D5 stacks and NS5 stacks. The integer $N$ is the D3 flux wrapping the annulus and is given as a function of the $\gamma_a,\delta_a,\hat\gamma_b,\hat\delta_b$ parameters and the annulus half-period $t$. We review the dictionary between gauge theory data and the supergravity parameters in Appendix \ref{app:FluxCFTDict}. One important point is that the triples $(\rho,\hat\rho,N)$ are defined up to certain shift ambiguities associated to large gauge transformations which affect the D3 fluxes (see \eqref{HWmove1},\eqref{HWmove2}). This phenomenon is related to the Hanany--Witten D3-brane creation/annihilation effect in the flat brane setup, which arises as one moves five-branes around the $x^3$ circle.

Explicitly the supergravity solution associated to the SCFT $C^\rho_{\hat\rho}(SU(\ti N),N)$ is given by the real harmonic functions on the annulus $\Sigma$,
\bea
\label{h1h2general}
h_1(z,\bar z)  &= -  \sum_{a=1}^p \gamma_a \ln\left[ \frac{\vartheta_{1}\left( \frac{\pi}{2} + i (z - \delta_a)  \big| \frac{it}{\pi} \right)}{\vartheta_{2}\left( \frac{\pi}{2} + i (z - \delta_a)  \big| \frac{it}{\pi}  \right)} \right] + c.c. \,,\cr
h_2(z,\bar z)  &= -  \sum_{b=1}^{\hat p} \hat\gamma_b \ln \left[ \frac{\vartheta_{1}\left(-i(z -\hat\delta_b) \big| \frac{it}{\pi} \right)}{\vartheta_{2}\left(-i(z -\hat\delta_b) \big| \frac{it}{\pi} \right)} \right] + c.c.  \,,
\eea
where the real parameters $\delta_a,\hat\delta_b$ are defined up to an overall real shift. $p$ and $\hat p$ are the number of D5 and NS5 stacks respectively. The parameters $\gamma_a$ and $\hat\gamma_b$ are positive and equal respectively to the number of D5-branes in the stack $a$ and to the number of NS5-branes in the stack $b$ (in the convention $\alpha'=4$).

\begin{figure}[h!]
\centering
\includegraphics[scale=0.75]{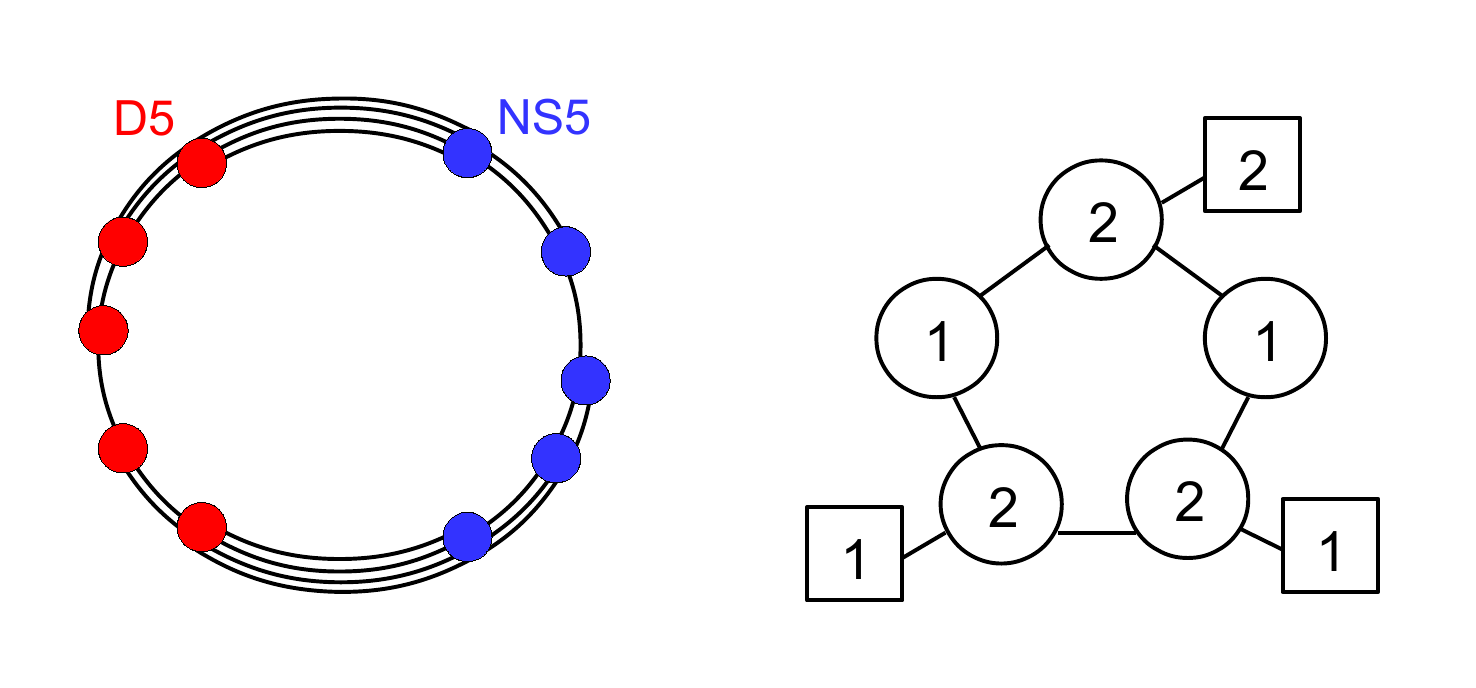} 
\vskip -0.5cm
\caption{\footnotesize A brane realization of a circular quiver invariant under the action of $\cS$ (translation + S-duality). The gauge theory data are read from the linking numbers of the five-branes: $\rho = (2,1,-1,-1,-1)$, $\hat\rho = (1,1,1,-1,-2)$. $N$ is the number of D3s stretched between the D5s and NS5s at the top, here $N=4$. On the right is the associated good circular quiver.}
\label{SinvQuiver}
\end{figure}

Circular quivers invariant under $\cS$ can be characterized by the fact that there exists a certain gauge\footnote{This specific gauge is not the one chosen in \cite{Assel:2012cj} to express the constraints $\rho^T \ge \hat\rho$ and $N>0$. In this other gauge $\rho,\hat\rho$ are really partitions, namely they are sets of positive integers, which is not the case here.} where the partitions $\rho,\hat\rho$ are of the form
\bea
\rho &= (\ell_1,\ell_2,\cdots,\ell_k) \,, \quad \ell_1 \ge \ell_2 \ge \cdots \ge \ell_k \,, \cr
\hat\rho &= (-\ell_k,\cdots, -\ell_2,-\ell_1) := -\rho \,.
\label{SdualPartitions} 
\eea
In this gauge the partitions sum to zero, $\sum_i \ell_i = 0$, meaning that the total D3 flux emanating from the D5 or NS5 singularities is vanishing. An example of a brane configuration realizing such a circular quiver is shown in Figure \ref{SinvQuiver}.
The condition \eqref{SdualPartitions} can be re-expressed as 
\be
\rho = (\rho_1, -\rho_2) \,, \quad \hat\rho = (\rho_2, -\rho_1) \,,
\label{SdualPartitions2}
\ee
with $\rho_1,\rho_2$ two partitions of an integer $M\ge 0$ with positive or zero coefficients. In the example of Figure \ref{SinvQuiver} these partitions are $\rho_1 = (2,1)$, $\rho_2 = (1,1,1)$.
Note that this is the criterion for $\cS$ invariance, but it is not the criterion for the usual S invariance implemented by the $S'$ action \eqref{Sprimeaction}, which corresponds to the condition that there exists a gauge where $\rho=\hat\rho$. Both criteria lead to gauge theories which are self-dual under mirror symmetry.\footnote{The half-ABJM theory discussed in Section \ref{ssec:HalfABJM} is an example of a theory with a gravity solution invariant under both types of S actions. This is not a generic example.} Here we are concerned only with quivers with partitions of the form \eqref{SdualPartitions}.

\begin{figure}[h!]
\centering
\includegraphics[scale=1]{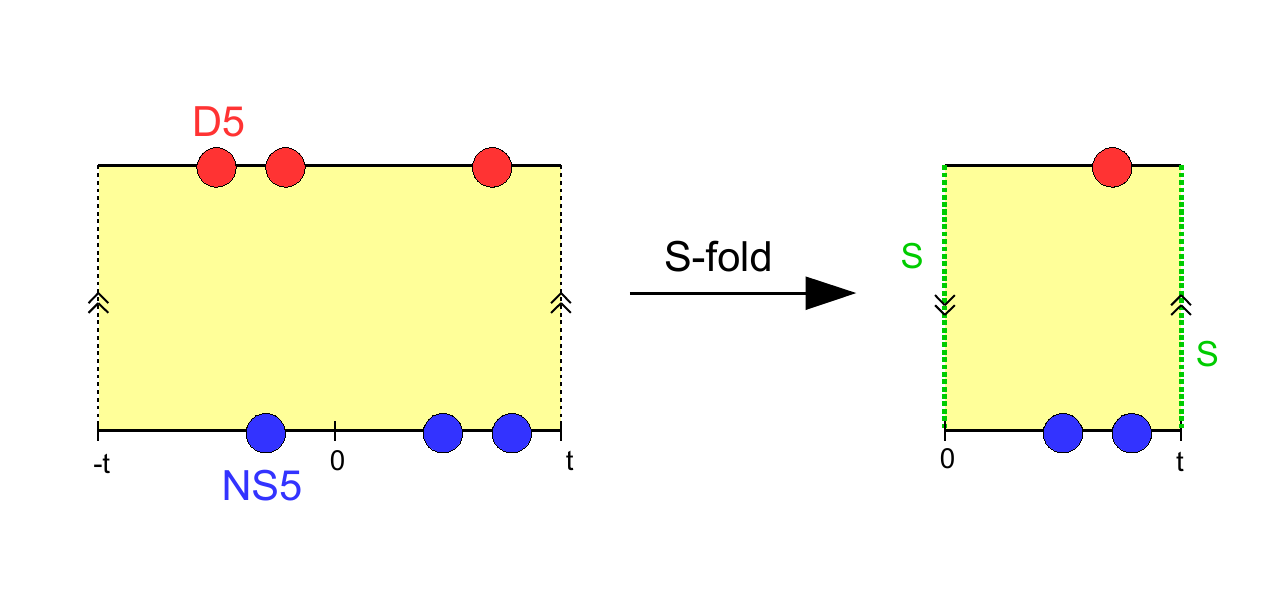} 
\vskip -1cm
\caption{\footnotesize An example of $\cS$-folding. The resulting S-flip solution has an S monodromy cut (green dashed line).}
\label{Sfolding2}
\end{figure}

Consistently with our discussion we find that the criterion \eqref{SdualPartitions} implies that the supergravity solution is invariant under the $\cS$ action \eqref{Saction}. In particular this means $p=\hat p$ and $(\gamma_a,\delta_a)=(\hat\gamma_a,\hat\delta_a+t \, \text{mod} \, 2t)$, for $a=1,\cdots, p$, where it is understood that $a$ labels the D5 stacks and the NS5 stacks from left to right along the $x$ axis.\footnote{This is not the same convention as in \cite{Assel:2012cj}.}
An $\cS$-invariant solution is therefore of the form
\bea
\label{h1h2Sinv}
h_1(z,\bar z) &= -  \sum_{a=1}^p \gamma_a \ln\left[ \frac{\vartheta_{1}\left( \frac{\pi}{2} + i (z - \delta_a)  \big| \frac{it}{\pi} \right)}{\vartheta_{2}\left( \frac{\pi}{2} + i (z - \delta_a)  \big| \frac{it}{\pi}  \right)} \right] + c.c. \,,\cr
h_2(z,\bar z)  &= -  \sum_{a=1}^{p} \gamma_a \ln \left[ \frac{\vartheta_{1}\left(-i(z -\delta_a -t) \big| \frac{it}{\pi} \right)}{\vartheta_{2}\left(-i(z -\delta_a - t) \big| \frac{it}{\pi} \right)} \right] + c.c.  \,.
\eea
The solution quotiented by $\cS$ in \eqref{Saction} is then described by the same harmonic functions $(h_1,h_2)$ restricted (for instance) to the domain $0\le x < t$, with an S-interface at $x = t \sim 0$, which combines the actions of S-duality, $S^2$ exchange and $y$ reflection. The resulting topology of $\Sigma$ is again that of the M\"obius strip. Here as well the quotient preserves only 3d $\N=3$ supersymmetry. An example of a quotient by $\cS$ is depicted in Figure \ref{Sfolding2}.

\begin{figure}[h!]
\centering
\includegraphics[scale=0.65]{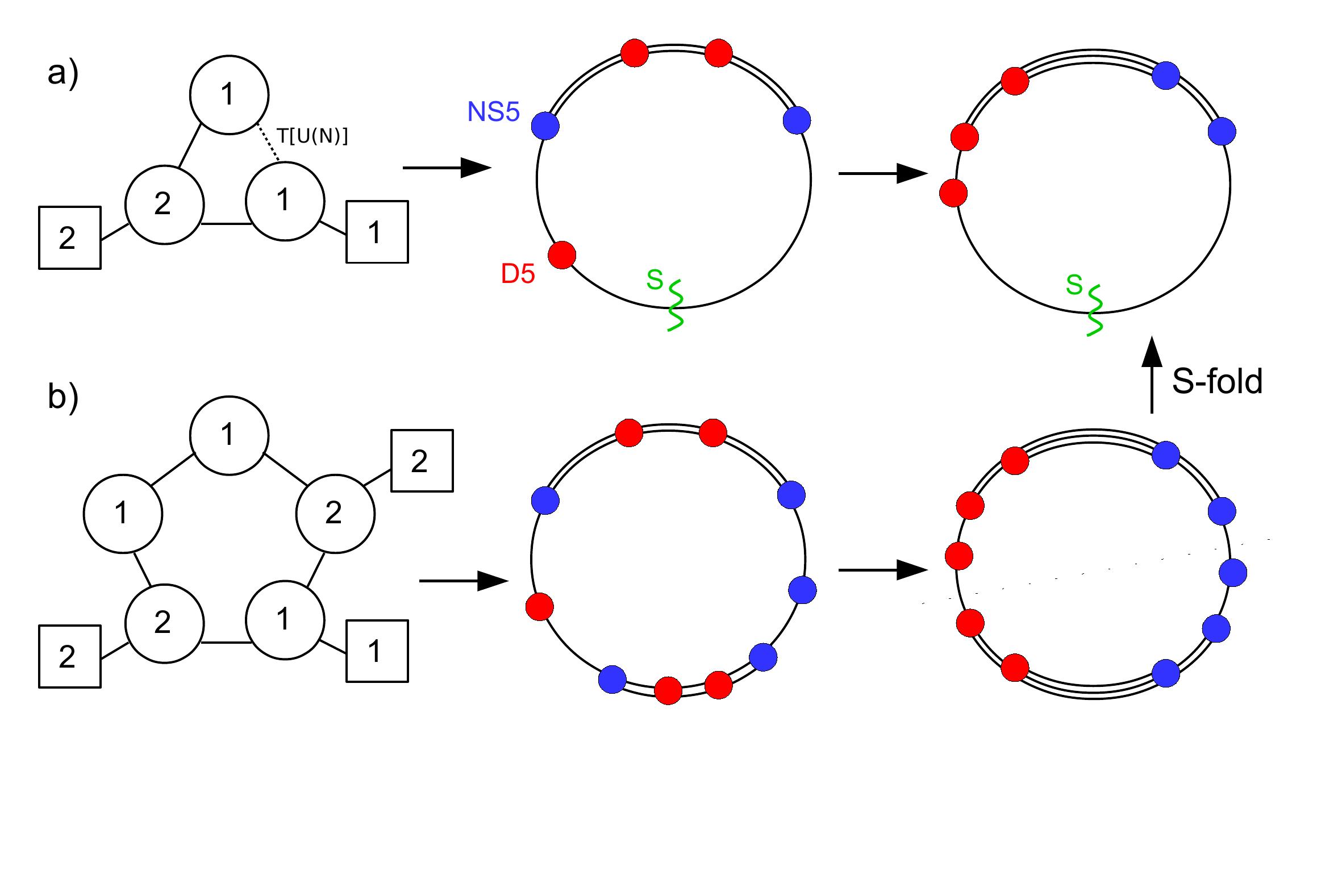} 
\vskip -2cm
\caption{\footnotesize a) An S-flip quiver and its brane realization. On the right: a canonical brane realization after Hanany--Witten moves, giving the quiver data $\rho_1 = (1,1,0)$, $\rho_2=(1,1)$, $N=3$ (number of D3s at the top). b) The parent circular quiver and its brane realization. On the right: the same brane configuration after Hanany--Witten moves, and the quotient by $\cS$.}
\label{SfoldEx}
\end{figure}

If the S-flip quiver is realized with $n$ D5s and $m$ NS5s, it can be constructed as the $\cS$-quotient of a circular quiver with $n+m$ D5s and $n+m$ NS5s. In Figure \ref{SfoldEx}-a we show an example of an S-flip quiver theory and we describe the associated brane configuration with an S-interface. We also show a canonical rearrangement of the five-branes from which one can read the two partitions $\rho_1,\rho_2$.
Such an S-flip quiver can be obtained as the $\cS$-quotient of a good circular quiver theory and the corresponding $\cS$-fold supergravity solution can be constructed from the solution of the `parent' quiver theory. In Figure \ref{SfoldEx}-b we describe the parent circular quiver with $\cS$ invariance, its brane configuration and the same brane configuration after Hanany--Witten moves with separated five-branes.

This construction applies to any S-flip quiver which is a good circular quiver with a $T[U(N')]$ link connecting two gauge nodes of minimal rank $N'$ in the quiver. We obtain a holographic map for a class of S-flip quivers labeled by two partitions with positive or zero coefficients and a positive integer, $(\rho_1,\rho_2, N')$, or alternatively by a zero-sum partition $\rho = (\rho_1,-\rho_2)$ (i.e. an array of positive and negative integers summing to zero) and an integer $N$, with $N=N'+M$ and $M = \sum_i \rho_{1,i} = \sum_i \rho_{2,i}$. 
Reciprocally, it is not hard to show that any good circular quiver invariant under $\cS$ gives rise after choosing the $\cS$-quotient appropriately  to an S-flip theory with the $T[U(N')]$ link connecting nodes of minimal rank. 

The S-flip solutions appear naturally parametrized by two partitions $\rho_1,\rho_2$ and an integer $N'$. However not all partitions describe an S-flip quiver theory. To ensure that the gauge node ranks in the quiver description are positive, one should take $N'>0$ and restrict to partitions satisfying the Young tableaux inequalities\footnote{This follows from a reasoning identical to that of \cite{Assel:2012cj} leading the Young tableau inequality $\rho^T \ge \rho$ in that paper.}
\be
\rho_1^T \ge \rho_2 \,.
\label{Constraints}
\ee
We thus obtain a holographic dictionary for a class of 3d $\N=3$ S-flip theories labeled by $(\rho_1,\rho_2,N')$ (or $(\rho,N)$) satisfying the constraints \eqref{Constraints}.\footnote{There is a redundancy in the parametrization with $\rho_1$ and $\rho_2$ because of the zeros in the partitions which can be transferred from $\rho_1$ to $\rho_2$ and vice-versa without changing the theory.} 
As a holographic check one should find that these constraints are satisfied by the partitions written in terms of the supergravity data.
 It is not very hard to find that these inequalities are implied by the inequalities $\rho^T \ge \hat\rho$ of the parent good circular quiver. Since the holographic map is consistent with these inequalities for the parent theories, we deduce that the holographic map is also consistent for the S-flip theories.

\subsection{Large $N$ free energy and holographic test}
\label{ssec:FreeEnergy}

In this section we perform a holographic test by computing the large $N$ three-sphere free energy of S-flip theories and comparing it to the regularized on-shell action of the dual gravity solution. As we shall see, a difficulty arises in the computation because the free energy is not easy to study in the regime of parameters where the supergravity approximation is valid. We will present only partial results here, postponing to future work a more complete analysis.

\subsubsection{Free energy}
\label{sssec:Free energy}

The matrix model computing the exact three-sphere partition function of the S-flip quiver theories are found using the rules described in Appendix \ref{app:ZS3}. In particular it includes a factor $Z_{T[U(N)]}(\sigma, \ti\sigma)$ \eqref{TUNFactor} for the $T[U(N)]$ link in the quiver. 
For instance for the half-ABJM theory of Section \ref{ssec:HalfABJM} we obtain, after simplifications, the matrix model
\be \label{eq:half-ABJM}
Z_{\rm half-ABJM} = \frac{1}{N!} \int d^N\sigma d^N\ti\sigma \,  \frac{\prod_{i<j}\sh(\sigma_{ij}) \sh(\ti\sigma_{ij})}{\prod_{i,j}\ch(\sigma_i-\ti\sigma_j)} \, e^{2\pi i \sum_i\sigma_i\ti\sigma_i}  \,.
\ee
The matrix model for the mirror of the half-ABJM theory is instead
\be
Z_{\rm half-ABJM \, mirror} = \frac{1}{N!} \sum_{\tau\in S^N} (-1)^\tau  \int d^N\sigma \frac{e^{2\pi i \sum_i \sigma_i \sigma_{\tau(i)}}}{\prod_i \ch \, \sigma_i}   \,.
\ee
Since the two theories are dual, their sphere partition function should be equal (up to finite counter-term ambiguities). This is easy to verify by using the Cauchy determinant formula
\be
\frac{\prod_{i<j}\sh(\sigma_{ij}) \sh(\ti\sigma_{ij})}{\prod_{i,j}\ch(\sigma_i-\ti\sigma_j)} = \sum_{\tau \in S^N} (-1)^\tau \frac{1}{\prod_{i=1}^N \ch(\sigma_i - \ti\sigma_{\tau(i)})} 
\ee
in the half-ABJM matrix model and then integrating over the $\sigma_i$ eigenvalues, with $\int dx \frac{e^{2\pi i x y}}{\ch(x-z)} =  \frac{e^{2\pi i zy}}{\ch \, y}$. This match is already a consistency check of our holographic construction.
\medskip

\begin{figure}[h!]
\centering
\includegraphics[scale=0.8]{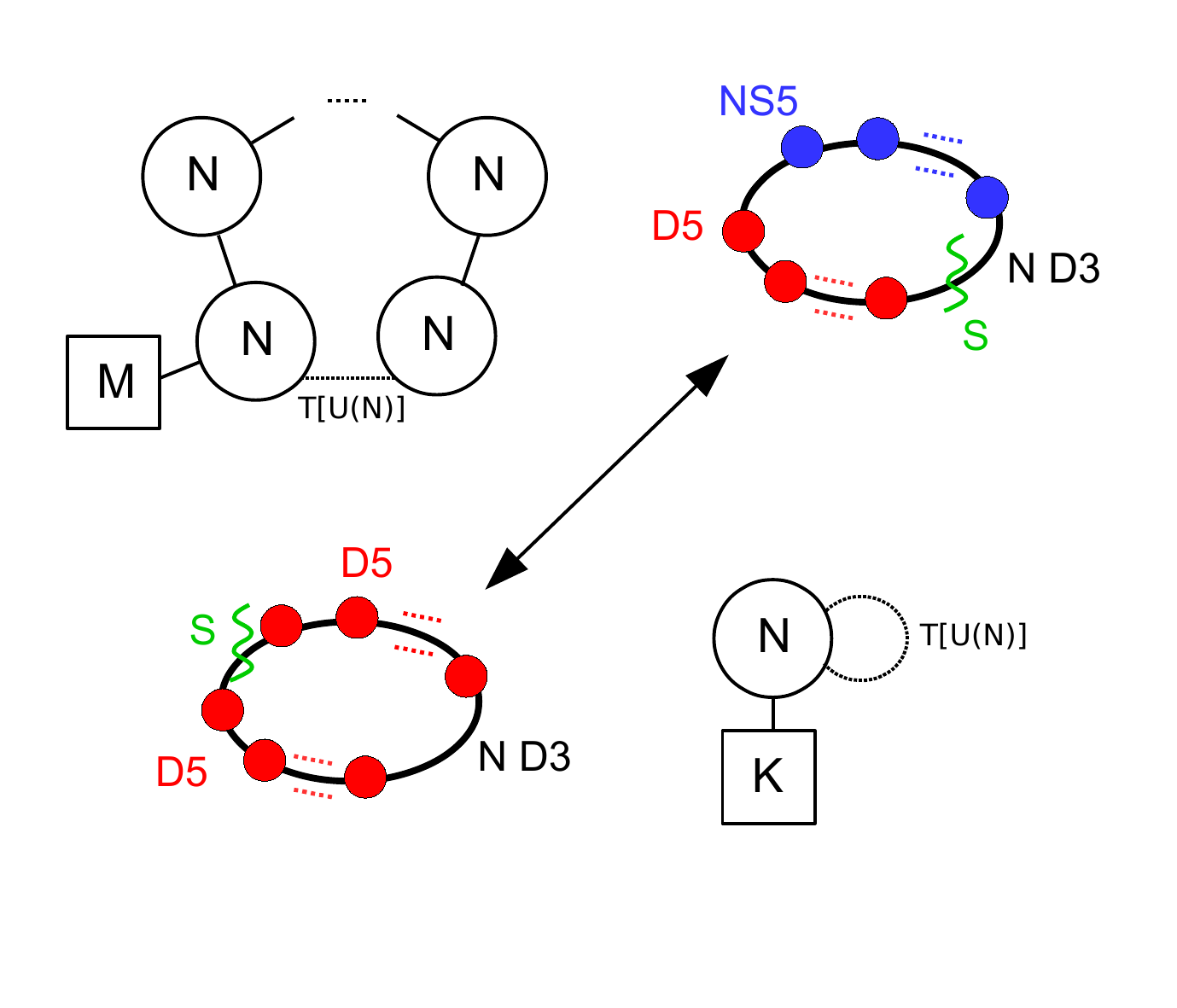} 
\vskip -1cm
\caption{\footnotesize Quiver with gauge group $U(N)^{\wat M+1}$, $M$ fundamental hypermultiplets and a $T[U(N)]$ link. The corresponding brane realization has $\wat M$ NS5s, $M$ D5s and an $S$ interface. After moving the $S$ interface we can reach a description with $K=M+\wat M$ D5s, corresponding to a simpler quiver theory (pure D5 dual theory).}
\label{Quiver2}
\end{figure}

We now generalize to an S-flip quiver theory with  $\wat M +1$ gauge nodes and with $M$ fundamental hypermultiplets distributed in various nodes.
When all the gauge node ranks are of the same order $N$ much larger than the differences between these ranks, we expect that the leading term in the free energy is only sensitive to $N$, to the total number of gauge nodes $\wat M +1$, and to the total number of fundamental hypermultiplets $M$, thus it is enough for our purposes to consider the circular quivers with gauge group $U(N)^{\wat M+1}$, with $M$ fundamental hypermultiplets in one node and with a $T[U(N)]$ link between two nodes, as shown in the upper part of Figure \ref{Quiver2}. The corresponding brane realization has $\wat M$ NS5-branes, $M$ D5-branes and an $S$ interface. The zero-sum partition describing such a quiver is simply a collection of $M+\wat M$ zeros.

The description of the theory can even be further simplified by going to a dual description obtained by moving the $S$ interface in the brane realization along the $x^3$ circle until it reduces to $\wat M + M$ D5-branes and an $S$ interface, as in the lower part of Figure \ref{Quiver2}. The corresponding dual theory has a single gauge node $U(N)$ with $K \equiv \wat M +M$ fundamental hypermultiplets and a self-$T[U(N)]$ link. We will refer to this alternative field theory description as the ``pure D5" dual description.\footnote{Of course, there are many other dual descriptions obtained by moving the $S$ interface at different positions along $x^3$. They form an orbit of ``mirror dual" theories.} From this argument we understand that the large $N$ free energy should only depend on $N$ and $K$. 

As a non-trivial test of this proposal we can show that the sphere partition functions of the dual theories match. This computation was already done in \cite{Assel:2014awa} and we reproduce it in Appendix \ref{app:PartFuncManipulations}. Since the three-sphere partition function of the initial theory and that of the pure D5 theories are equal, we can use the latter to study the large $N$ free energy. The matrix model of the pure D5 theory reads
\be
Z = \frac{1}{N!} \sum_{\tau\in S^N} (-1)^\tau \int d^N\sigma  \frac{e^{2\pi i \sigma.\sigma_\tau}}{\prod_i\ch( \sigma_i)^{K}}  \,.
\label{ZSfoldMM}
\ee
We are not able to evaluate this seemingly simple matrix model, however we can study more easily its large $K$ limit.
Taking $K$ large, the leading contribution comes from the region $\sigma_j \sim \frac{1}{\sqrt K}$ and the matrix model can be approximated by
\be
Z \simeq \frac{1}{N!} \sum_{\tau\in S^N} (-1)^\tau \int d^N\sigma  \frac{1}{2^{KN}} e^{2\pi i \sigma.\sigma_\tau} e^{-\frac{\pi^2}{2} K \sum_{j=1}^N \sigma_j^2}  \,.
\ee
This is essentially the same matrix model as for the partition function of $J_n$ theories (up to complex conjugation), which we evaluated in Appendix \ref{app:JnPartFunc}. We find (up to a phase)
\be
Z = \frac{e^{\frac{NT'}{2}}}{2^{KN}\prod_{j=1}^N \left((-1)^j e^{j T'} -1\right) }  \,,
\ee
with $e^{T'} + e^{-T'} = \frac{i\pi K}{2}$, $e^{T'} = \frac{i\pi K}{4} \lp 1 + \sqrt{1 + \frac{16}{\pi^2 K^2}} \rp$.
This leads to the free energy
\be
 F = KN \ln 2 + \frac 12 N^2 \ln K + O(K^0) \,,
 \label{Fnaive}
 \ee
 at large $K$. In this derivation we have kept $N$ finite and taken $K$ large. However, to compare with the supergravity on-shell action we will need to assume large $N$ and it is not clear whether our evaluation holds in this limit. Let us look in some more detail at the approximation above. We can define $a_j = \pi \sqrt K \sigma_j$ and expand the $\ch(...)$ functions ($\equiv 2\cosh(\pi(...))$) at large $K$ and fixed $a_j$,
 \bea
\frac{1}{\ch(\sigma_j)^K} &= \frac{1}{2^K\cosh(\pi\sigma_j)^K} = \frac{1}{2^K\cosh(a_j/\sqrt{K})^K} = e^{-\frac{a_j^2}{2} + \frac{a_j^4}{12 K} + O(\frac{1}{K^2})}  \cr
&= e^{-\frac{a_j^2}{2}}\left(1 + \frac{a_j^4}{12K} + O\left(\frac{1}{K^2}\right)\right) \,. 
\eea
Plugging the expansion in $Z$ we get
\bea
Z &= \frac{1}{N!} \sum_{\tau\in S^N} (-1)^\tau \int d^Na \, \frac{e^{\frac{2i}{\pi K} \sum_i a_i a_{\tau(i)}}}{\pi^N K^{N/2}2^{KN}}\,  e^{-\frac{1}{2} \sum_{i=1}^N a_i^2} \left(1 + \frac{1}{12K} \sum_{i=1}^N a_i^4 + O\left(\frac{1}{K^2}\right)\right) \,.
\eea
The approximation that we did consists in dropping the terms in the parenthesis after the 1. The next term after 1 is $\frac{1}{12K} \sum_{i=1}^N a_i^4$ and its contribution to $Z$ scales with a factor $\frac{N}{K}$ compared to the first contribution, therefore it is subleading only if $K \gg N$. This indicates that our approximation is valid only when $K \gg N$. 
Since we are interested in the free energy which is the logarithm of $Z$, we can even trust our approximation for $F$ when the correction is of the same order as the leading term (since it only corrects $F$ by a constant). We conclude that our approximation is valid when $N \lesssim K $,
\be
\underline{ K \gg 1 \  \text{and} \  N \lesssim K}: \qquad   F = KN \ln 2 + \frac 12 N^2 \ln K \,.
 \label{Fnaive2}
 \ee
We notice that when $K \gg N$, the first term $KN \ln 2$ dominates, while for $K \sim N \gg 1$ the second term $\frac 12 N^2 \ln K$ dominates, so there seems to be a phase transition at $K \sim N$.

\subsubsection{On-shell action and holography}
\label{sssec:OSaction}

\begin{figure}[h!]
\centering
\includegraphics[scale=0.8]{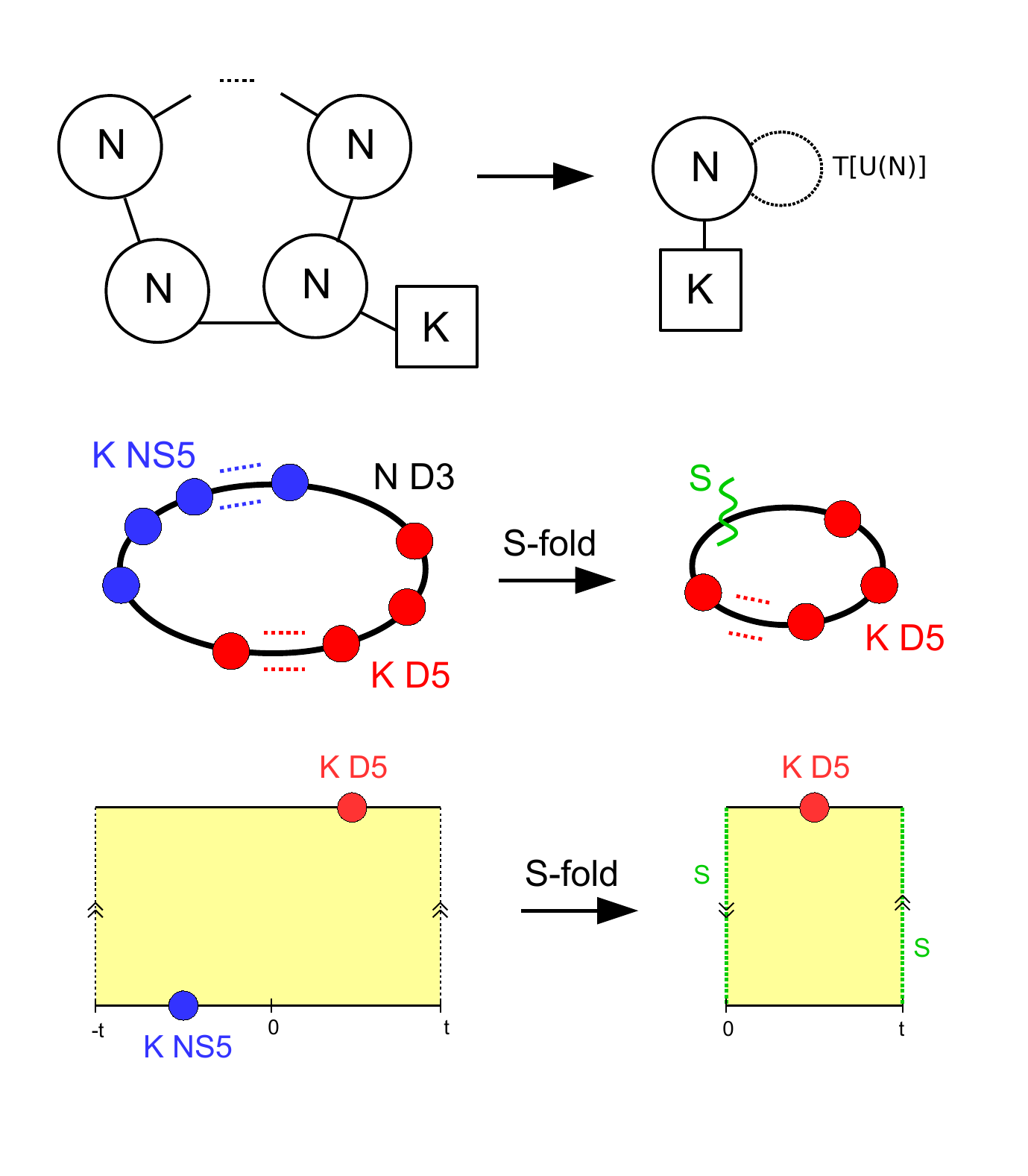} 
\vskip -0.5cm
\caption{\footnotesize $\cS$-quotient leading to the pure D5 theory. The double cover theory has $K$ $U(N)$ nodes and $K$ fundamental hypermultiplets in one node. The double cover supergravity solution has single stacks of $K$ D5s and $K$ NS5s.}
\label{Quiver2Sfold}
\end{figure}

The SCFT free energy must be compared with the large $N$ on-shell action of the dual supergravity solution. This solution is very similar to the half ABJM solution. It has a single stack of D5-branes (or a single stack of NS5-branes in another S-duality frame). It is obtained by starting from the ``double cover" circular quiver theory with one stack of $K$ D5s and one stack of $K$ NS5s (this corresponds to $\rho = \hat\rho = (0,0, \cdots, 0)$ with $K$ zeros), and doing the quotient by $\cS$. The operation on the brane setups and the corresponding supergravity solutions are shown in Figure \ref{Quiver2Sfold}.

The supergravity solution is the same as for the half-ABJM theory, with harmonic functions multiplied by $K$,
\bea
\label{h1h2Quiver2}
h_1(z,\bar z) &= -  K \ln\left[ \frac{\vartheta_{1}\left(\frac{\pi - i t}{2} + i z  \big| \frac{it}{\pi} \right)}{\vartheta_{2}\left( \frac{\pi - i t}{2} + i z  \big| \frac{it}{\pi}  \right)} \right] + c.c. \,,\cr
h_2(z,\bar z) &= -  K\ln \left[ \frac{\vartheta_{1}\left(-i(z +\frac t2) \big| \frac{it}{\pi} \right)}{\vartheta_{2}\left(-i(z +\frac t2) \big| \frac{it}{\pi} \right)} \right] + c.c.  \,,
\eea
on the M\"obius strip $(x,y)\sim (x+t,\frac{\pi}{2} - y)$ with $S$ duality gluing conditions.
The parameter $t$ is now given in terms of $N$ by the relation \eqref{NtRel}, which has an extra factor of $K^2$ due the presence of the five-brane stacks \cite{Assel:2012cj},
\bea
N &= \frac{2}{\pi} K^2 \sum_{n\ge 0} (2n+1)\, \text{arctan}\lp e^{-2(n + 1/2)t}\rp  \,.
\label{NtRelQuiver}
\eea
The regularized on-shell action is evaluated as before with using the formula \eqref{OnShellAction}.
There are two natural limits one can study: small $t$ or large $t$.
\medskip

\noindent {\bf Small $t$ limit (or fat M\"obius strip)}: 

In the small $t$ limit we have
\be
N = \frac{K^2}{2t^2} \int_0^{+\infty} du \, u\frac{2}{\pi} \text{arctan}(e^{-u}) + O(t^0) = \frac{\pi^2}{32 t^2}K^2 + O(t^0)  \,.
\ee
It corresponds in terms of field theory data to having $N \gg K^2$,
\be \label{eq:N>>K2}
t \ll 1 \quad \leftrightarrow \quad N \gg K^2 \,.
\ee
In this limit the five-brane stacks gets smeared along the upper and lower boundaries of $\Sigma$ (see \cite{Assel:2012cj}).
Using the asymptotics of Jacobi theta functions one finds
\bea
h_1(z,\bar z) &=  -\frac{i\pi K}{2t} z + c.c. + O(t) \,, \cr
h_2(z,\bar z) &=  \frac{i\pi K}{2t} \lp z - \frac{i\pi}{2}\rp + c.c. + O(t) \,,
\eea
leading to
\be  \label{eq:Ssmallt}
S_{IIB} = \frac{\pi^4}{24}\frac{K^4}{(2t)^3} = \frac{\pi}{3\sqrt{2}} K N^{\frac 32} \,,
\ee
at leading order at large $N$.

In this small $t$ limit, it is not clear whether higher derivative corrections to the supergravity action may not be neglected. The dilaton goes from $\infty$ at $y=0$ to $-\infty$ at $y=\pi/2$. The curvature $R_{(s)}$ and dilaton factors $(\nabla\phi)^2$ are suppressed by $1/\sqrt{N}$ factors but they are functions of $y$ over $\Sigma$ and have both divergences at $y=0$ and $y=\frac{\pi}{2}$, which are the location of the smeared five-branes. Therefore these corrections are not small in the full geometry and it is not clear whether the supergravity computation above is valid or not. Moreover the contribution of the D5-brane action could also compete with $S_{IIB}$.
\medskip

\noindent{\bf Long wavelength limit (or long M\"obius strip)}:  

The alternative limit $t \gg 1$ may be more appropriate to the holographic test. In this limit the fields of the supergravity solution vary slowly with $x$ in the bulk geometry. In terms of field theory data we now have
\be
N = \frac{2}{\pi}K^2 e^{-t} \,,
\ee
so the large $t$ limit corresponds to $N \ll K^2$,
\be
t \gg 1 \quad \leftrightarrow \quad  N \ll K^2 \,. 
\ee
 In this limit the harmonic functions reduce to
\bea
\underline{0 < x < \frac t2 :} \quad   h_1(z) &= -2 i K e^{-\frac t2 + z} + c.c. \,, \cr
h_2(z) &=  2 K e^{-\frac t2 - z} + c.c. \,, \cr
\underline{\frac t2 < x < t :} \quad   h_1(z) &= 2 i K e^{\frac t2 - z} + c.c. \,, \cr
h_2(z) &=  2 K e^{-\frac{3t}{2} + z} + c.c. \,. \cr
\eea
This is the same local geometry as in the extremal Janus solution with appropriate factor identification.

To study the validity of the supergravity approximation it is enough to focus on the region with $0\le x \le \frac t2$, since the region $\frac{t}{2} \le x \le t$ is related to it by a reflection.
For the higher derivative corrections to the IIB supergravity action to be suppressed we require that $R_{(s)}$ and $g_{(s)}^{\mu\nu}\nabla_\mu \phi\nabla_\nu \phi$ be small, as discussed in Section \ref{ssec:TestJanus}.

In the region $0<x<\frac t2$ we have $g_{\mu\nu} \sim K e^{-t/2} \sim \sqrt{N}$ and  $e^{2\phi} \sim e^{-2x}$. 
We find $g_{(s)}^{\mu\nu}\nabla_\mu \phi\nabla_\nu \phi  \sim e^{x}/\sqrt{N}$ and $R_{(s)} \sim e^{x}/\sqrt{N}$. 
The maximal values are obtained at $x = t/2$ and are of order $e^{t/2}/\sqrt{N} \sim K/N$.

Putting things together we find that the supergravity approximation is a priori valid in the regime
\be
K \lesssim N \ll K^2 \,,
  \label{NKlimits}
\ee
where we allowed the limiting case $K \sim N$ for which $R_{(s)}$ and $(\nabla \phi)^2$ become order one only in a thin region around $x=t$. The range of validity can be recast as $N^{1/2} \ll K \lesssim N$. We also assume that the contributions from the five-brane effective actions is subleading in this limit.
The evaluation of the on-shell action proceeds as before and gives at leading order at large $t$,
\be
S = \frac{2}{\pi^2} K^4 e^{-2t} t = \frac{1}{2} N^2 \ln\lp\frac{K^2}{N} \rp \,.
\label{Slarget}
\ee

\bigskip

\noindent{\bf Holographic match}

We can now compare the result with our evaluation of the free energy \eqref{Fnaive2} for the dual CFT for ranges of $N$ and $K$ for which both computations are reliable. The range of validity of the two computations are almost non-overlapping and indeed the two evaluations are not the same. However there is still a borderline situation where both computations should be reliable and this is when $K$ and $N$ are of the same order, $K \sim N$. In that case both computations agree with 
\be
\underline{ K \sim N \gg 1}: \qquad   F = \frac 12 N^2 \ln N  \,. 
 \label{Fmatch}
 \ee
This is quite a non-trivial match, confirming the proposed holographic duality.

\bigskip 

\noindent{\bf Beyond the supergravity regime}

As we commented after (\ref{eq:Ssmallt}), in the small $t$ regime the backreaction of the branes is felt all over the geometry, and it is not clear that the supergravity approximation is valid. One might be curious to see what happens to the field theory computation in this regime, which as we saw in (\ref{eq:N>>K2}) corresponds to $N\gg K^2$. 

As an extreme test, we can take $K=1$, which corresponds to the original half-ABJM theory, whose matrix model we wrote in (\ref{eq:half-ABJM}). Its saddle configurations have $\sigma=\tilde \sigma$; a numerical study reveals that both the real and imaginary parts of $\sigma$ scale with $N^{1/3}$ in the large $N$ limit, and that they follow a linear pattern in the complex plane.
One can study this more precisely with the techniques of \cite{Herzog:2010hf,Jafferis:2011zi}, rewriting the matrix model in terms of a local action for a single continuum field $\sigma(x)$. This local action turns out to be formally identical to \cite[(8.2)]{Jafferis:2011zi}, and leads to $F_{K=1}\sim \frac{(3\pi)^{5/3}}{10} e^{-i\pi/6}N^{5/3}$. This $N^{5/3}$ scaling is typical of massive IIA solutions \cite{Aharony:2010af}. Its appearance in our IIB context is intriguing, and might be suggestive of a IIA dual in this limit. T-duality is not obviously applicable, because not all the fields are $x$-independent, but one can speculate that our solution is the back-reacted form of an object that does have a nontrivial IIA T-dual. (T-duality of non-perturbative objects is sometimes challenging in supergravity, and often requires some smearing.)


\section*{Acknowledgements}

We thank Costas Bachas, Yuji Tachikawa and Alberto Zaffaroni for fruitful discussions at various stages of the project. A.T.~is supported in part by INFN and by the ERC Starting Grant 637844-HBQFTNCER.


\appendix

\section{Local supergravity solutions and $SL(2,\bR)$ action}
\label{app:LocalSol}

The general local solutions of IIB supergravity with $OSp(4|4)$ invariance were found in \cite{D'Hoker:2007xy}. They are parametrized by two harmonic functions $h_1,h_2$ on a Riemann surface $\Sigma$. One first introduces auxiliary functions
\be
W = \p\bar\p(h_1 h_2) \,, \quad N_j = 2h_1h_2|\p h_j|^2 - h_j^2 W \,.
\ee
The Einstein frame metric is given by
\be
ds^2 = f_4^{\, 2} ds^2_{\rm AdS_4} + f_1^{\, 2} ds^2_{{\rm S}^2_1} + f_2^{\, 2} ds^2_{{\rm S}^2_2} + 4\rho^2 dz d\bar z \,,
\ee
with warp factors
\bea
f_4^{\, 8} &= 16 \frac{N_1 N_2}{W^2} \,, \quad  
f_1^{\, 8} &= 16 h_{1}^8 \frac{N_{2} W^2}{N_{1}^3} \,, \quad 
f_2^{\, 8} = 16 h_{2}^8 \frac{N_{1} W^2}{N_{2}^3} \,, \quad 
\rho^8 = \frac{N_1 N_2 W^2}{h_1^4 h_2^4}  \,.
\eea
The solution is written in an $SL(2,\bR)$ gauge where the axion field vanishes. The axio-dilaton takes the (purely imaginary) form
\be
\tau = \chi + i e^{-2 \phi} = i \sqrt{N_1 \over  N_2}\,.
\ee
Note that we use an unconventional normalization for the dilaton $\phi$.
In addition there are  3-form and 5-form backgrounds. To specify the corresponding gauge potentials we need to introduce dual harmonic functions $h_1^D, h_2^D$,
\bea
h_1 = -i (\cA_1 - \bar\cA_1) \quad &\rightarrow \qquad h_1^D = \cA_1 + \bar\cA_1 \,, \cr
h_2 = \cA_2 + \bar\cA_2 \quad &\rightarrow \qquad h_2^D = i (\cA_2 - \bar\cA_2) \,.
\eea
The dual harmonic functions are defined up to constant shifts corresponding to large gauge transformations of the background.
The NS-NS and R-R 3-forms are expressed by
\be
H_3  =   \omega^{45}\wedge db_1  \,, \qquad  F_3 =   \omega^{67}\wedge db_2   \,,
\ee
where $ \omega^{45}$ and $ \omega^{67}$ are the volume forms of the unit-radius  spheres  ${\rm S}_1^2$ and ${\rm S}_2^2$, and
\bea
b_1 &= 2 i h_1 \frac{h_1 h_2 (\p  h_1\bar  \p  h_2 -\bar \p  h_1 \p  h_2)}{N_1} + 2  h_2^D \,,  \cr
b_2 &= 2 i h_2 \frac{h_1 h_2 (\p  h_1 \bar\p  h_2 - \bar\p  h_1 \p  h_2)}{N_2} - 2  h_1^D \,.
\eea
The expression for the self-dual 5-form is a little more involved and we refer to the original papers \cite{D'Hoker:2007xy,D'Hoker:2007xz} for its expression.

The choice of harmonic functions $h_1,h_2$ is constrained by a number of conditions ensuring the regularity of the solution. In particular the boundary of $\Sigma$ must be divided into regions where one harmonic function obeys (vanishing) Dirichlet boundary conditions while the other obeys Neumann boundary conditions. 
\medskip

Other solutions can be generated by $SL(2,\bR)$ transformations of the above background, which act on the axio-dilaton and 3-form fields as follows,
\be
\tau' = \frac{d\tau+c}{b\tau+a} \,, \qquad
\lp   
\begin{array}{c}
H'_3  \\
F'_3
\end{array}
\rp =
\lp
\begin{array}{cc}
 a & b \\ 
 c & d
\end{array}
\rp
\lp   
\begin{array}{c}
H_3  \\
F_3
\end{array}
\rp \,.
\ee
The Einstein-frame metric and the 5-form are invariant under $SL(2,\bR)$.

With this choice of conventions, the $SL(2,\bZ)$ subgroup is generated by the elements
\be
S =\lp\begin{array}{cc} 0 & -1 \cr 1 & 0 \end{array}\rp \,, \quad  T =\lp\begin{array}{cc} 1 & 0 \cr 1 & 1 \end{array}\rp \,,
\ee
satisfying $S^2 = -1$ and $(ST)^3 =1$, and acting on $\tau$ by $S.\tau = -1/\tau$, $T.\tau = \tau +1$.

\section{Sphere partition function and matrix models}
\label{app:ZS3}

The (undeformed) three-sphere partition function $Z$ of a 3d $\N \ge 2$ supersymmetric Lagrangian gauge theory can be computed exactly via supersymmetric localization, as first shown in \cite{Kapustin:2009kz} following the seminal work of Pestun \cite{Pestun:2007rz}. The final evaluation of the sphere partition function is expressed as a matrix model,\footnote{Here and later we suppress pure phase factors of the matrix model which play no role in the computation of the large $N$ free energy.}
\be
Z = \int \frac{d\sigma}{|W|} \, Z_{\rm vec}(\sigma) \, Z_{\rm chiral}(\sigma) \, Z_{\rm CS}(\sigma) \, Z_{\rm FI}(\sigma) \,,
\ee
where $\sigma = \{\sigma_i\}$ are the (real) "eigenvalues" taking values in the Cartan subalgebra of the gauge group, $|W|$ is the order of the Weyl group, and the integrand is a product of contributions associated to the vector multiplet, chiral multiplets, Chern--Simons terms and FI terms. The factors appearing in the integrand  simplify for an $\N\ge 3$ theory. For a $U(N)$ gauge group with fundamental matter we have $N$ eigenvalues $\sigma_{i=1,\cdots,N}$, $|W|=N!$ and the vector multiplet and fundamental hypermultiplet factors
\bea
& Z^{\N=4}_{\rm vec}(\sigma) = \prod_{i<j=1}^N \sh^2(\sigma_{ij}) \,, \qquad   Z_{\rm hyper}(\sigma) = \prod_{i=1}^N  \frac{1}{\ch(\sigma_i - m )} \,,
\label{VecFactor}
\eea
with the notation $\sh(x) \equiv 2\sinh(\pi x)$, $\ch(x) \equiv 2 \cosh(\pi x)$ and $\sigma_{ij}\equiv \sigma_i - \sigma_j$. The parameter $m$ above is a real mass for the hypermultiplet.
The factor for a bifundamental hypermultiplet of $U(N)\times U(\ti N)$ is
\be
Z_{\rm hyper}(\sigma,\ti\sigma) =  \prod_{i=1}^N \prod_{j=1}^{\ti N}  \frac{1}{\ch(\sigma_i -\ti\sigma_j - m )} \,.
\label{HyperFactor}
\ee
Finally the contributions of a supersymmetric $U(N)$ Chern--Simons term at level $k\in\bZ$ and an FI term with parameter $\eta$ are
\be
Z_{\rm CS}(\sigma) = e^{i\pi k \sum_{i=1}^N \sigma_i^2} \,, \qquad  Z_{\rm FI}(\sigma) = e^{2i\pi \eta \sum_{i=1}^N \sigma_i} \,.
\label{CSFactor}
\ee
The coupling to a $T[U(N)]$ theory by gauging its $U(N)\times U(N)$ global symmetry is encoded in a contribution $Z_{T[U(N)]}(\sigma,\ti\sigma)$ to the matrix integrand, where $\sigma_{i=1,\cdots,N}, \ti\sigma_{i=1,\cdots,N}$ are the eigenvalues of the two $U(N)$ symmetries. This contribution is nothing but the sphere partition function of the $T[U(N)]$ theory which was computed via supersymmetric localization in \cite{Benvenuti:2011ga}. It is given by
 \be
Z_{T[U(N)]}(\sigma,\ti\sigma) = \frac{\sum_{\tau\in S_N} (-1)^\tau e^{2\pi i \sum_{i=1}^N \sigma_i\ti\sigma_{\tau(i)}}}{\prod_{i<j}\sh(\sigma_{ij}) \prod_{i<j}\sh(\ti\sigma_{ij})} \,.
\label{TUNFactor}
\ee
The identification of this factor as describing an $S$ duality interface contribution was first described in \cite{Gulotta:2011si} and further studied in \cite{Assel:2014awa}. 

We explained in section \ref{ssec:JnfoldSCFTs} that there are two possible ways to couple the $T[U(N)]$ theory, by gauging $U(N)\times U(N)$ or $U(N)\times U(N)^{\dagger}$. One gauging corresponds to inserting the above factor in the matrix model, while the other gauging corresponds to inserting $Z_{T[U(N)]}(\sigma,-\ti\sigma)$. The relevant gauging for the $J_n$ theory is associated to the insertion of $Z_{T[U(N)]}(\sigma,-\sigma)$ in the matrix model as in \eqref{ZJnMM}.

\section{Sphere partition function of $J_n$ theories}
\label{app:JnPartFunc}

In this appendix we evaluate the matrix model \eqref{ZJnMM} computing the partition function $Z(N)$. We will be able to compute it axactly, namely at fintie $N$. For computational convenience we add an overall phase $e^{-\frac{i\pi N}{4}}$ to the matrix model.

The matrix model (with the extra phase) has the form of a Fermi gaz partition function \cite{Marino:2011eh}
\be
Z(N) = \frac{1}{N!}\sum_{\tau\in S_N} (-1)^\tau \int d^N\sigma \, \prod_{i=1}^N \bra{\sigma_i} \hat\rho \ket{\sigma_{\tau(i)}} \,, 
\ee
with density operator
\be
\bra{\ti\sigma} \hat\rho \ket{\sigma} = e^{-\frac{i\pi}{4}} \, e^{i\pi n \ti\sigma^2} \, e^{-2\pi i \ti\sigma\sigma} := \rho(\ti\sigma,\sigma) \,. 
\ee
It is convenient to change ensemble and to define the grand canonical partition function
\be
\Theta(z) = 1 + \sum_{N\ge 1} Z(N) z^{N} := e^{J(\mu)} \,, \quad z := e^{\mu}  \,.
\ee
The grand canonical potential $J(\mu)$ then  takes the form 
\be
J(\mu) = -\sum_{\ell\ge 1} \frac{(-1)^{\ell}}{\ell} Z_{\ell} \, e^{\mu\ell}\,,
\ee
with 
\bea \label{eq:Zell}
Z_\ell &= \tr(\hat\rho^\ell) = \int d\sigma \bra{\sigma} \hat\rho^\ell \ket{\sigma} = \int d^\ell\sigma\,  \rho(\sigma_\ell,\sigma_1) \prod_{a=1}^{\ell-1} \rho(\sigma_a,\sigma_{a+1}) \cr
& =  e^{-\frac{i\pi \ell}{4}} \int d^\ell\sigma \, e^{i\pi n \sigma_\ell^2} e^{-2\pi i \sigma_\ell\sigma_1} \prod_{a=1}^{\ell-1} e^{i\pi n \sigma_a^2} e^{-2\pi i \sigma_a\sigma_{a+1}} \,.
\eea
The evaluation of $Z(N)$ can be extracted from $e^{J(\mu)}$ by an inverse Legendre transform and the problem is reduced to computing $Z_\ell$ (which is simpler that $Z(N)$) and performing the inverse Legendre transform.

To evaluate $Z_{\ell}$, we  write 
\be \label{eq:zcell}
Z_\ell = e^{-i\pi \ell/4}\int d^\ell \sigma \exp[i\pi \sigma^t C_{\ell,n} \sigma] = (\det C_{\ell,n})^{-1/2} \,,
\ee
where $\sigma = (\sigma_1,\ldots, \sigma_\ell)^t$ and the matrix
\begin{equation}\label{eq:Cell}
	C_{\ell,n}= n 1_\ell  - P - P^t\,,
\end{equation}
with $P$ is the matrix realizing the permutation $(23\ldots \ell1)$. $C_\ell(n)$ is a cyclic matrix. The eigenvalues of $C_{\ell,n}$ are $n - \omega_i - \omega_i^{-1}$, where $\omega_i$ are the $\ell$-th roots of unity, $\omega_i^\ell=1$. Hence its determinant is
\be
\det C_{\ell,n} = \prod_{i=0}^{\ell-1} (n - \omega_i - \omega_i^{-1}) \,.
\ee 
Now consider the meromorphic function 
\be
f(z)\equiv \prod_{i=0}^{\ell-1} (z + z^{-1}+ \omega_i + \omega_i^{-1})= e^{i\pi(1-\ell)}\prod_{i=0}^{\ell-1}(1+z\omega_i)(1+z^{-1}\omega_i) \,.
\ee
It is ${\Bbb Z}_\ell$-invariant: $f( \omega_i z)= f(z)$. Hence its Laurent series can only contain integer powers of $z^\ell$. From its definition as a product it is clear that it can in fact only contain $z^\ell$, $z^{-\ell}$ and a constant, and in fact that $f(z)= z^\ell + z^{-\ell}+ f_0$, with $f_0$ a constant. To fix $f_0$, consider the case $z=-1$. We have $f(-1)= (-)^\ell\det C_{\ell,-2}$; but $C_{\ell,-2}$ is minus the Cartan matrix of an affine Lie algebra, which is known to have zero determinant. Hence we have $0= f(-1)= 2(-1)^{\ell}+f_0$, leading to $f(z)= z^\ell + z^{-\ell} - 2 (-1)^\ell$. Now notice that, if we define $T$ such that $n=e^T+e^{-T}$ as in (\ref{JnParam}),
\begin{equation}\label{eq:detcl}
	\det C_{\ell,n}= (-1)^\ell f(-e^T) = e^{\ell T} + e^{-\ell T} - 2 = (e^{\ell T/2} - e^{-\ell T/2})^2\,.
\end{equation}
From (\ref{eq:zcell}) it now follows
\be \label{eq:Zellres}
Z_\ell = \frac{1}{e^{\frac{\ell T}{2}} - e^{-\frac{\ell T}{2}} } \,.
\ee
From here the grand potential $J(\mu)$ can be computed as follows
\bea
J(\mu) &= -\sum_{\ell\ge 1} \frac{(-1)^{\ell}}{\ell} \frac{1}{e^{\frac{\ell T}{2}} - e^{-\frac{\ell T}{2}} } \, e^{\mu\ell} \cr
&= -\sum_{\ell\ge 1} \sum_{m\ge 0} \frac{(-1)^{\ell}}{\ell}  e^{-\ell T (m+1/2)} \, e^{\mu\ell}  \cr
&= \sum_{m\ge 0} \ln \left( 1 +  e^{\mu - T (m+1/2)} \right) \,,
\eea
leading to the grand canonical partition function
\be
\Theta(z) = \prod_{m\ge 0} \left( 1 +  z \, e^{- T (m+1/2)} \right) = (-z \, e^{- T/2}; e^{-T})_\infty \,,
\ee
where we used the Pochhammer symbol in the last expression.
The partition function $Z(N)$ is recovered by the residue computation
\bea
Z(N) &= \int \frac{dz}{2\pi i z^{N+1}} \Theta(z)  = \int \frac{dz}{2\pi i z^{N+1}} \prod_{m\ge 0} \left( 1 +  z \, e^{- T (m+1/2)} \right) \cr
&= \sum_{m_1 > m_2 > \cdots > m_N\ge 0} e^{-NT/2} e^{-T\sum_{i=1}^N m_i} \,.
\eea
The sum over $m_i$ are simple geometric series which can be performed one by one leading to the remarkably simple final expression
\be
Z(N) = \frac{ e^{\frac{NT}{2}}}{\prod_{j=1}^N \left( e^{j T} - 1 \right) } \,.
\label{ZJnFinalApp}
\ee


\section{Fluxes in elliptic solutions and quiver data}
\label{app:FluxCFTDict}

In this appendix we briefly review the dictionary between the triples $(\rho,\hat\rho,N)$ describing good circular quiver theories and the data $(\gamma_a,\delta_a,\hat\gamma_b,\hat\delta_b,t)$ of the elliptic supergravity solutions dual to their fixed point SCFTs, as described in \cite{Assel:2012cj}.

The triple $(\rho,\hat\rho,N)$ is obtained by considering the brane realization of the circular quiver, with D3s, D5s and NS5s, and by moving the five-branes along the circle direction $x^3$ until all D5s are on one side and all NS5s are on the other side. Because of Hanany--Witten brane creation effects one obtains a configuration with various D3s ending on both types of five-branes. The net number of D3s\footnote{For D5s, this is the number of D3s ending on its left minus the number of D3s ending on its right. For NS5s it is the opposite number.} ending on a given five-brane is called its linking number. The partition $\rho$ is the array of D5s linking numbers, the partition $\hat\rho$ is the array of NS5s linking numbers and $N$ is the total number of D3-branes at a chosen position in $x^3$.
 There is large redundancy in the choice of triples $(\rho,\hat\rho,N)$ which describe a given good circular quiver. The holographic dictionary proposed in \cite{Assel:2012cj} uses a `gauge' with
\bea
\rho &= (\ell_1,\ell_2,\cdots,\ell_k) \,, \quad \ell_1 \ge \ell_2 \ge \cdots \ge \ell_k \ge 0 \,, \cr
\hat\rho &=  (\hat\ell_1,\hat\ell_2,\cdots,\hat\ell_{\hat k}) \,, \quad \hat\ell_1 \ge \hat\ell_2 \ge \cdots \ge \hat\ell_{\hat k} \ge 0 \,,
\eea
satisfying the constraint $\rho^T \ge \hat\rho$. 
To compare with the supergravity data, one should re-label the partitions in terms of `stacks',
\bea
\rho &= (\underbrace{\ell^{(1)}, \cdots, \ell^{(1)}}_{N^{(1)}},\cdots,\underbrace{\ell^{(p)}, \cdots, \ell^{(p)}}_{N^{(p)}}) \,, \cr
\hat\rho &=  (\underbrace{\hat\ell^{(1)}, \cdots, \hat\ell^{(1)}}_{\hat N^{(1)}},\cdots,\underbrace{\hat\ell^{(p)}, \cdots, \hat\ell^{(p)}}_{\hat N^{(\hat p)}}) \,.
\eea
The number $\gamma_a$, resp. $\hat\gamma_b$ of D5-branes, resp. NS5-branes, in a stack is given by
\bea
\gamma_a &= N^{(a)} \,, \quad a=1,\cdots, p \,, \cr
\hat\gamma_b &= \hat N^{(b)} \,, \quad b =1, \cdots,\hat p \,.
\eea
The linking numbers are mapped to the D3 flux emanating from a given five-brane stack, averaged over the number of five-branes in the satck. They are related to the data of the supergravity solution by 
\bea
\ell^{(a)}  &=  \sum_{b=1}^{\hat p} \hat N^{(b)}  \left[ \sum_{n=0}^{\infty} f(\hat\delta_b - \delta_a -2nt) - \sum_{n=1}^{\infty} f(-\hat\delta_b + \delta_a - 2nt)
 \right] \,, \cr
\hat\ell^{(b)} &=  -\sum_{a=1}^p N^{(a)}\left[ \sum_{n=1}^{\infty} f(-\hat\delta_b + \delta_a - 2nt)
- \sum_{n=0}^{\infty} f(\hat\delta_b - \delta_a - 2nt) 
 \right] \,.
 \eea
with $f(x) = \frac{2}{\pi} \arctan(e^{x})$.
Finally the paramter $N$ is identified with the D3 flux going through the annulus with
\be 
N = \sum_{a=1}^p \sum_{b=1}^{\hat p} N^{(a)} \hat N^{(b)}  \sum_{n=1}^{\infty} n  \left[ f(\hat\delta_b - \delta_a - 2nt)
 + f(\delta_a - \hat\delta_b - 2nt) \right] \,.
\ee

Changes of gauge affect the above formulas. They correspond to moving five-brane stacks around the $x^3$ circle in the flat brane picture. The basics moves are
\bea
(\ell_1,\ell_2,\cdots,\ell_k) \quad &\to \quad  (\ell_2, \cdots, \ell_k, \ell_1-\hat k) \cr
(\hat\ell_1,\hat\ell_2,\cdots,\hat\ell_{\hat k})  \quad &\to \quad (\hat\ell_1-1,\hat\ell_2-1,\cdots,\hat\ell_{\hat k}-1) \cr
N \quad &\to \quad  N + \hat k -\ell_1 \,,
\label{HWmove1}
\eea
and 
\bea
(\ell_1,\ell_2,\cdots,\ell_k) \quad &\to \quad  (\ell_1-1,\ell_2-1,\cdots,\ell_k-1) \cr
(\hat\ell_1,\hat\ell_2,\cdots,\hat\ell_{\hat k})  \quad &\to \quad (\hat\ell_2,\cdots,\hat\ell_{\hat k},\hat\ell_1 -k) \cr
N \quad &\to \quad  N + k -\hat\ell_1 \,,
\label{HWmove2}
\eea
as well as the reverse moves.


\section{Equivalence of $S^3$ partition functions}
\label{app:PartFuncManipulations}

The matrix model computing the three-sphere partition function of the initial theory $U(N)^{\wat M+1}$ with $M$ fundamental in one node is, after some simplification,
\bea
Z &= \int \prod_{a=1}^{\wat M+1} \left[ \frac{d^N\sigma^{(a)}}{N!} \prod_{i<j}\sh^2(\sigma^{(a)}_{ij}) \right] \prod_{a=1}^{\wat M}\frac{1}{\prod_{i,j}\ch(\sigma^{(a)}_i-\sigma^{(a+1)}_j)} \cr
& \phantom{= \int } \frac{1}{\prod_i \ch(\sigma^{(1)}_i)^M} \frac{N! e^{2\pi i \sigma^{(\wat M +1)}.\sigma^{(1)}}}{\prod_{i<j}\sh(\sigma^{(\wat M+1)}_{ij}) \prod_{i<j}\sh(\sigma^{(1)}_{ij})} \,,\eea
with the notation $\sigma_{ij} = \sigma_i - \sigma_j$, $\sigma.\ti\sigma = \sum_{i=1}^N \sigma_i \ti\sigma_i$.
The standard trick to simplify the matrix model expression is to use the Cauchy identity
\be
\frac{\prod_{i<j}\sh(\sigma_{ij})\prod_{i<j}\sh(\ti\sigma_{ij})}{\prod_{ij}\ch(\sigma_i - \ti\sigma_j)} = \sum_{\rho\in S^N}\frac{(-1)^\rho}{\prod_i \ch(\sigma_i - \ti\sigma_{\rho(i)})} \,.
\ee
After simplifications we end up with
\bea
Z &= \int \prod_{a=1}^{\wat M+1} d^N\sigma^{(a)} \frac{1}{N!} \sum_{\rho\in S^N} (-1)^\rho \frac{e^{2\pi i \sigma^{(\wat M +1)}.\sigma^{(1)}_\rho}}{\prod_i\ch(\sigma^{(1)}_i)^M}  \prod_{a=1}^{\wat M}\frac{1}{\prod_{i}\ch(\sigma^{(a)}_i-\sigma^{(a+1)}_i)} \,.
\eea
Using the identity
\be
\frac{1}{\ch(y)} = \int dx \frac{e^{2\pi i x y}}{\ch(x)}  \,,
\ee
one can reach the following form of the matrix model
\be
Z = \int d^N\sigma \frac{1}{N!} \sum_{\rho\in S^N} (-1)^\rho \frac{e^{2\pi i \sigma.\sigma_\rho}}{\prod_i \ch(\sigma_i)^{M+\wat M}}  \,,
\ee
corresponding to the matrix model of the pure D5 dual theory.


\bibliography{Benbib}
\bibliographystyle{JHEP}

\end{document}